\documentclass[12pt]{iopart} 

\usepackage{iopams,amsfonts,amssymb,amsthm} 

\pdfoutput=1
\usepackage{etex}
\usepackage{cite}
\usepackage{latexsym}
\usepackage{rawfonts}
\usepackage{graphicx}
\usepackage[usenames,dvipsnames,svgnames]{xcolor}
\usepackage{bbm}
\usepackage{latexsym}
\usepackage{multirow}
\usepackage{rotating}
\usepackage{lscape}
\usepackage{graphicx} 
\usepackage{subfigure}
\usepackage{xcolor}
\usepackage{fancybox}
\usepackage{ifmtarg}

\input prepictex
\input pictex
\input postpictex

\usepackage{cmbright}
\usepackage[T1]{fontenc}
\usepackage{graphicx}

\def\q{\quad}
\def\qq{\qquad}
\def\qqq{\qquad\qquad}

\def\2;{\;\;}

\def\eps{\epsilon}

\def\IntN{{\mathbb Z}}

\def\mathL{{\mathbb L}}

\def\shalf{{\sfrac{1}{2}}}

\def\e#1{{\vec{e}}_#1}

\def\o#1{\overline{#1}}

\def\Ref#1{(\ref{#1})}

\def\C#1{{\mathcal #1}}

\def\y#1{{y_#1}}

\def\binom#1#2{{{#1}\choose{#2}}}
\def\Bi#1#2{{\binom{#1}{#2}}}

\def\SBi#1#2{\hbox{\footnotesize ${\binom{#1}{#2}}$}}

\def\Sfrac#1#2{\hbox{\large $\frac{#1}{#2}$}}
\def\sfrac#1#2{\hbox{\nor $\frac{#1}{#2}$}}

\def\LB{\left(}         \def\RB{\right)}
        
\def\LA{\left\langle}        \def\RA{\right\rangle}

\def\lfl{\!\left\lfloor} \def\rfl{\right\rfloor\!}
       
\def\LH{\left[}        \def\RH{\right]}

\def\nor{\normalsize}

\def\vv{{\;\hbox{\Large $|$}\;}}

\def\edge#1#2{{\langle #1 \hspace{0.85pt}{\sim}\hspace{0.85pt} #2 \rangle}}
\def\dedge#1#2{{\langle #1\hspace{1.25pt} {\leadsto}\hspace{0.50pt} #2\rangle}}

\def\ande{\:\hbox{and}\;}

\def\thin{ {\hspace{0.75pt}} }
\def\plus{{\hspace{0.85pt}{+}\hspace{0.85pt}}}
\def\minus{{\hspace{0.85pt}{-}\hspace{0.85pt}}}

\usepackage{cmbright}
\usepackage{textcomp}
\usepackage[font=footnotesize,labelfont=sf]{caption}

\definecolor{blue}{rgb}{0,0.18,0.39}
\definecolor{RoyalBlue}{rgb}{0,0.2,0.7}

\begin{document}
\title{Forces and pressures in adsorbing partially directed walks}
\author{
E.J. Janse van Rensburg$^1$\footnote[1]{\texttt{rensburg@yorku.ca}}
and T. Prellberg$^2$\footnote[2]{\texttt{t.prellberg@qmul.ac.uk}}
}

\address{$^1$Department of Mathematics and Statistics, 
York University, Toronto, Ontario M3J~1P3, Canada\\}

\address{$^2$School of Mathematical Sciences,
Queen Mary, University of London,
Mile End Road, London E1~4NS, United Kingdom}

\begin{abstract}
Polymers in confined spaces lose conformational entropy.  This induces a net
repulsive entropic force on the walls of the confining space.  A model for
this phenomenon is a lattice walk between confining walls, and in this
paper a model of an adsorbing partially directed walk is used. 
The walk is placed in a half square lattice 
$\mathL^2_+$ with boundary $\partial \mathL^2_+$,
and confined between two vertical parallel walls, which are vertical lines in 
the lattice, a distance $w$ apart. The free energy of the walk is determined, 
as a function of $w$, for walks with endpoints in the confining walls 
and adsorbing in $\partial \mathL^2_+$.  This gives the entropic force 
on the confining walls as a function of $w$.  It is shown that there are 
zero force points in this model and the locations of these points are 
determined, in some cases exactly, and in other cases asymptotically.
\end{abstract}

\maketitle

\section{Introduction}
\label{section1}   

A linear polymer in a good solvent placed near a geometric obstacle (such as a hard wall) 
loses conformational entropy.  This loss of entropy induces a net force on the obstacle
\cite{MN91}, and if the polymer can move freely, then it will tend to move away from the 
obstacle.  The induced force is repulsive, and the polymer exerts an average net 
pressure on the wall.   These entropic pressures have been seen
in experiments \cite{BDD95,CS95,GWF06} and have been modelled numerically 
(see for example reference \cite{JD13}).   

A similar situation is seen when a polymer is placed in a confined space; the loss in
conformational entropy induces a net repulsive force on the walls of the confined
space.  For example, a polymer between two colloidal particles loses entropy as
the particles approach one another;  this induces a net repulsive force between
the two particles.  The repulsion between particles is the mechanism underlying
the stabilization of a colloid by a polymer \cite{NDA97}.  This phenomenon was 
examined numerically using a self-avoiding walk model of a polymer confined
between two hard walls \cite{DR71}.  A self-interaction self-avoiding walk
model of polymer pulled at its endpoints was similarly considered in reference
\cite{JD13,Guttmann09}. See references \cite{SSW88A,WS91,WS92} 
for more results.

In this paper a two dimensional partially directed walk model is used to
model  the entropic forces induced by an adsorbing polymer between confining
walls.  In addition, the entropic pressure of an adsorbing polymer on the adsorbing 
wall will be examined as well.   Similar models were examined in reference
\cite{JvRP13}, using directed path models to model the pressure of a
directed path on the adsorbing boundary.  Related work on 
the entropic pressure near knotted lattice ring polymers was
reported in reference \cite{JvR14}, which built on the results in
references \cite{JvRR12,GJvR13}.

In this paper we continue the study of entropic pressure near lattice polymers
by using a partially directed walk model in the square lattice.   Directed path models
were examined in reference \cite{JvRP13}, where the entropic pressure near directed paths
and staircase polygons models (of a grafted linear polymer) was calculated.  In particular,
for a directed path of length $n$ from the origin in the positive half-lattice, the pressure
on the $x$-axis a distance proportional to $an$ from the origin is given by
\begin{equation}
P_n(a) = - \frac{2\sqrt{2}}{\sqrt{\pi\,n^3\,a^3(1\minus a)}} + O(n^{-5/2}) .
\end{equation}
This, in particular, shows that
\begin{equation}
\lim_{n\to\infty} n^{3/2}P_n(a) = - \frac{2\sqrt{2}}{\sqrt{\pi\,a^3(1\minus a)}}
\end{equation}
as $n\to\infty$. 

Partially directed walk models of polymers were introduced
in references \cite{CP88,W98}, and have been widely used as models of polymer
entropy; see for example references \cite{F90,BORW05,BOR09}.  The generating
function of adsorbing partially directed walks has been computed in 
several models using the Temperley method \cite{T56}, and in this paper we
will follow a similar approach to that of reference \cite{OP10}.

Let $\IntN^2$ be the \textit{square grid} with standard basis $\{\e1,\e2\}$.
A point or \textit{vertex} $\vec{x}\in \IntN^2$ is a vector with Cartesian 
coordinates $(\vec{x}(1),\vec{x}(2))$.  Two points $\vec{x},\vec{y}\in\IntN^2$ 
are \textit{adjacent} in $\IntN^2$ if $\|\vec{x}-\vec{y}\|_2 = 1$. The \textit{edge} 
$\edge{\vec{x}}{\vec{y}}$ is a unit length line segment between adjacent 
points (or vertices) $\vec{x}$ and $\vec{y}$ in $\IntN^2$.  The 
\textit{square lattice} $\mathL^2$ is the set of all edges between
adjacent vertices of $\IntN^2$, defined by
\begin{equation}
\mathL^2 = \{ \edge{\vec{x}}{\vec{y}} \vv 
\hbox{$\vec{x},\vec{y}\in \IntN^2$ and $\|\vec{x} \minus  \vec{y}\|_2=1$}\} .
\end{equation}
The \textit{positive square lattice} is given by
\begin{equation}
\mathL^2_+ =  \{ \edge{\vec{x}}{\vec{y}} \in\mathL^2 \vv 
\hbox{$\vec{x}(2)\geq 0$ and $\vec{y}(2)\geq 0$}\} .
\end{equation}
The \textit{boundary} of $\mathL^2_+$ is given by
\begin{equation}
\partial \mathL^2_+=  \{ \edge{\vec{x}}{\vec{y}} \in\mathL^2_+ \vv 
\hbox{$\vec{x}(2)= 0$ and $\vec{y}(2) =0$}\} 
\end{equation}
and every edge in $\partial \mathL^2_+$ is parallel to $\e1$ and normal
to $\e2$.

A \textit{directed edge} is a directed unit length line segment from a 
vertex $\vec{x}$ to a vertex $\vec{y}$ in $\IntN^2$.  It is denoted by
$\dedge{\vec{x}}{\vec{y}}$.  A directed edge $\dedge{\vec{x}}{\vec{y}}$
is in the East direction if $\vec{x}(1)\plus 1=\vec{y}(1)$ and $\vec{x}(2)=\vec{y}(2)$,
it is in the South direction if $\vec{x}(1)=\vec{y}(1)$ and $\vec{x}(2)\minus 1=\vec{y}(2)$
and it is in the North direction if $\vec{x}(1)=\vec{y}(1)$ and $\vec{x}(2)\plus 1=\vec{y}(2)$.

A \textit{partially directed walk} $\omega$ in $\mathL^2$ of length $n$ is a sequence of $n$ 
directed edges in the East (E), North (N) and South (S) directions, starting in $\vec{0}$, such
that a directed edge in the N direction cannot immediately be followed by a directed
edge in the S direction and vice versa.  That is, $\omega$ is a sequence of directed
edges in the E, N and S directions 
$\LA \dedge{\vec{v_0}}{\vec{v_1}},\dedge{\vec{v_1}}{\vec{v_2}},
\dedge{\vec{v_2}}{\vec{v_3}},\ldots,\dedge{\vec{v_{n-1}}}{\vec{v_n}}\RA$ such that 
$\vec{v}_0=\vec{0}$ and all the $\vec{v_j}$ are distinct (so that $\omega$ is
self-avoiding).   The terminal or last vertex of $\omega$ is $\vec{v}_n$.

If a directed walk from $\vec{0}$ is in the positive square lattice $\mathL^2_+$,
then it is a \textit{positive partially directed walk}.  A positive partially directed
walk is illustrated in figure \ref{Fig1}.

\begin{figure}[t!]
\centering
\input Fig1.tex
\caption{A partially directed walk from the origin in $\mathL^2_+$ of
width $w$ and endpoint at $height$ $h$.  The walk is weighted by
the activity $a$ conjugate to the number of edge-visits in 
$\partial\mathL^2_+$ and its endpoints are tethered to two vertical
walls a distance $w$ apart which bounds a confining space (vertical
slit of width $w$) containing the walk.  If $h=0$ then this is an
\textit{adsorbing bargraph path} from the origin in $\mathL^2_+$.}
\label{Fig1}   
\end{figure}

A partially directed walk from the origin in the positive half-lattice is illustrated in
figure \ref{Fig1}.  Edges in the walk which are in the $x$-axis are weighted by
the generating variable $a$.   For large $a$ the walk stays near the $x$-axis, so that
this is a model of an \textit{adsorbing partially directed walk}.  In section 
\ref{sectionW} the
generating function of walks of fixed width $w$ is determined using the kernel
method \cite{BM10,FS09}.  In particular, we show that the generating function of
partially directed walks from the origin in the half-lattice, of width $w$ with final 
vertex at height $h$, is
\begin{equation}
\fl
G_{wh}(a,y) =  
\sum_{s=0}^w
\sum_{k=0}^w 
\sum_{i\geq 0} \frac{w\plus 1\minus s}{w\plus i\plus 1\minus s}
\Bi{h}{w\minus k\minus s}
\Bi{n\minus 1}{i} \Bi{w}{s\minus i}\, \frac{y^{h+2i} (a\minus 1)^k  }{(1\minus y^2)^s} .
\label{eqnGwh-intro}   
\end{equation}
The variable $y$ generates vertical steps (and there are exactly $w$ horizontal steps).
This result can be used to extract the partition function of walks of length $n$, 
width $w$ and last vertex at height $h$ (see equation \Ref{eqn35}).

In section \ref{section4} the entropic forces in the model are examined in
the scaling limit, defined by putting $w=\lfl \alpha n \rfl$ for walks of length
$n$, and then taking $n\to\infty$ in the model (see figure \ref{Fig1}).  
This gives the limiting free energy $F (\alpha)$ of the walk, which is computed 
for several cases.  For example, for walks with $h=0$ and $a=1$, the limiting 
free energy is given by
\begin{eqnarray}
\fl F_0(\alpha) & \fl \qq\; = \alpha\log(2\alpha(1\minus\alpha)) 
-(\sqrt{C}\plus 2\alpha\minus 1)\log(\sqrt{C}\plus 2\alpha \minus 1) \nonumber \\
&\qq\qqq-(1\minus \sqrt{C})\log(1\minus\sqrt{C})
+\shalf\sqrt{C}\log \LB \Sfrac{C+\alpha}{C-\alpha}\RB\! ,
\end{eqnarray}
where $C=2\alpha^2-2\alpha+1$.
The derivative of this expression to $\alpha$ gives the limiting force
\begin{equation}
\fl
\C{F}_0(\alpha) = \sfrac{d}{d\alpha} F_0(\alpha) = 
\log \LB
\Sfrac{2\alpha(1-\alpha)}{(C\plus 2\alpha\minus 1)^2}
\RB
+ \Sfrac{C^2\minus 2\alpha}{2C}
\log \LB
\Sfrac{(C-\alpha)(2\alpha+C-1)^2}{(C+\alpha)(1-C)^2} \RB\! . \qq
\end{equation}
It follows directly from the above that $\C{F}_0(\alpha) 
= - \C{F}_0(1\minus \alpha)$ (so that $\C{F}_0(\sfrac{1}{2}) = 0$).  That is, the limiting 
force vanishes (in the limit) when the confining walls are a distance $w=\lfl\shalf n\rfl$
apart.  This is the \textit{zero force point} in the model, and it is located in this
model when the walk is extended one-half of its length along the adsorbing line
(or when one-half of the edges (steps) in the walk is in the horizontal direction).
The limiting force curve $\C{F}_0(\alpha)$ is also symmetric on reflection through the point 
$(\shalf,0)$ in the $(\alpha,\C{F}_0(\alpha))$-plane.  For small $\alpha>0$ the 
above shows that $\C{F}_0(\alpha) \sim |\log \alpha |$ (see equation \Ref{eqn63BB}).

In addition to the above, the asymptotic forces in other models are examined.
These models include walks with a lifted endpoint, as well as walks which are 
adsorbing.  In each case asymptotic expressions for the
limiting forces are determined, and the location of the zero force point is determined.

In section \ref{section5} finite size asymptotics of the model is developed 
for the case that $a=1$.  The partition function simplifies to a single summation 
which may be approximated using a saddle-point method.   We determine
an asymptotic expression for the partition function, and we use this
expression to determine the asymptotics for the repulsive force
between vertical walls (see section \ref{section4}).   In particular,
if the walk has length $n$ and the confining walls are a distance $\alpha n$ apart,
then the asymptotic force has leading term asymptotics $\C{F}_n(\alpha)
= - \sfrac{1}{\alpha} + (n\minus\sfrac{1}{2})  -n \log \alpha + O(n\alpha)$ 
as $\alpha \searrow 0$. For small values of $\alpha$ the force is positive
(repulsive).

\begin{figure}[t]
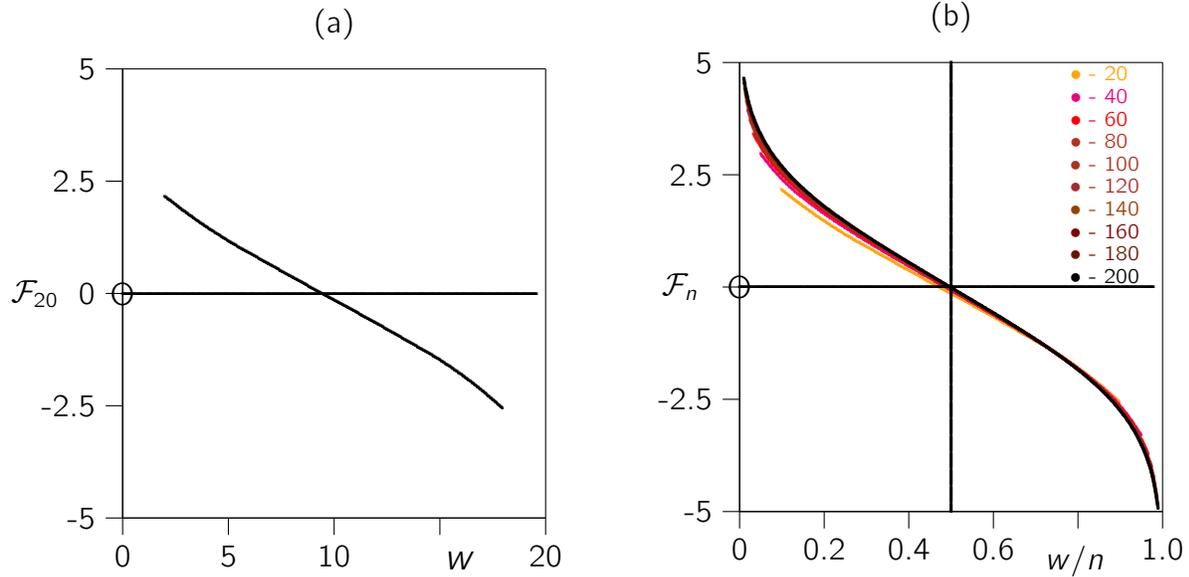

\input Fig2.tex
\caption{The entropic forces exerted by a partially directed walk 
confined between two vertical walls in $\mathL^2_+$
(as shown in figure \ref{Fig1}).  (a)  The force for walks of length $n=20$, endpoint
at height $h=0$, and with $a=1$, plotted as a function 
of $w$, the distance between the two walls (or equivalently, 
the width of the walk).  For small values of $w$ the force is positive (and therefore
repulsive, pushing the walls away from one another).  For large values of
$w$ the force is negative; this indicates an attractive force between the
two walls.  (b)  Force curves for walks of lengths from $n=20$ to
$n=200$, for $a=1$ and $h=0$, as a function of $w$.  
The horizontal length scale is normalised by $n$ for each curve (so that
$w=\lfl \alpha n\rfl$ for $0\leq \alpha \leq 1$); 
this collapses the curves into a single force curve shown which tends to
a limiting curve as $n\to\infty$. Notice that the forces vanish at approximately 
$w=\frac{1}{2} n$. }
\label{Fig2}   
\end{figure}

In section \ref{section6} the model is examined numerically.  This is done in particular
to (1) verify the asymptotic results, and (2) to examine cases which we have not
analysed in section \ref{section5}.   The approach is to determine
forces and pressures in the models, using as starting point the expressions for
the partition function and generating function determined in section \ref{sectionW}.
In the first instance the entropic forces  exerted on two confining vertical walls by
an adsorbing partially directed walk with endpoints tethered in the walls are determined
numerically (see figure \ref{Fig1}).  By rescaling the forces by the length of the walk,
the data collapse to a single force curve which is repulsive when the confining
walls are close together, and attractive when the confining walls are far apart.
The typical situation is seen in figure \ref{Fig2} when $a=1$ and the height of the
endpoint of the walk is $h=0$.  The force curves are modified when $a>1$, generally
becoming more repulsive as $a$ increases (and taking the walk through its transition
into its adsorbed phase).  

Secondly, the pressure of the adsorbing partially directed walks on the adsorbing wall 
is examined.  If the walk is tethered at the origin, then there is a large entropic
pressure on the adsorbing wall close to the origin.  This pressure decays quickly with
distance  from the origin.   If both endpoints of the walk are confined to the adsorbing
wall, then for walks of length $n$ a secondary pressure peak is seen at a distance
about $\sfrac{1}{2}n$ from the origin -- this peak is the result of the \textit{other}
endpoint of the walk (which exerts pressure in the vicinity of the point where
the path returns to the adsorbing wall; see for example figure \ref{Fig17}).

In section \ref{section7} we conclude the paper with a few final remarks.

\section{Partially directed walks of width $w$}
\label{sectionW}   

\begin{figure}
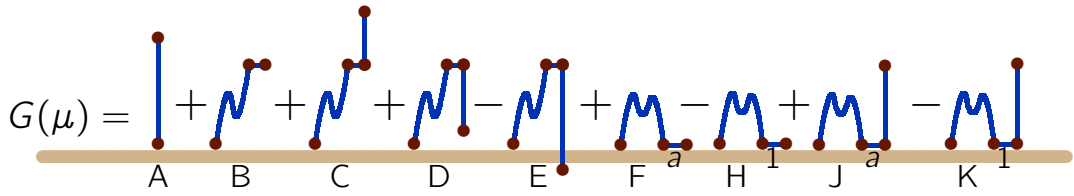

\input Fig5.tex
\caption{The generating function $G(\mu)$ of partially directed walks from the
origin in $\mathL^2_+$, with edge-visits weighted by $a$ and height of endpoint
weighted by $\mu$, may be classified as above.  This classification gives a 
functional recurrence for $G(\mu)$.  Each walk is either ($A$) a string of
vertical edges (possibly of length $0$), or ($B$) ends in a horizontal step at height
$h>0$, or ($C$) ends in a horizontal step followed by a non-empty sequence of steps
in the up direction, or ($D\minus E$) ends in a horizontal step followed by a non-empty
sequence steps in the South direction (downwards) which does not not step 
below $\partial\mathL^2_+$, or ($F\minus H$) ends in a horizontal step at 
height $h=0$ (substracted out by $H$ and added back in with weight $a$ 
in $F$), or ($J\minus K$),  ends in a horizontal edge at height $h=0$, followed by 
a non-empty sequence of vertical edges, subtracted out in $K$,
and then added back with weight $a$ in $J$. }
\label{Fig5}   
\end{figure}

In this section the kernel method (see for example reference 
\cite{BM10}) is used to determine $G(\mu) \equiv
G(x,y,\mu,a)$, the generating function of partially directed walks from the origin
in $\mathL^2_+$ with a collection of activities $\{ x,y,\mu,a\}$, conjugate to length, 
height of endpoint, and number of edge-visits to $\partial\mathL^2_+$. 
Introduce the following generating variables:
\begin{itemize}
\item horizontal steps: $x$;
\item vertical steps: $y$;
\item edge-visits to the adsorbing boundary: $a$;
\item height of last vertex: $\mu$.
\end{itemize}
The variable $\mu$ is the designated \textit{catalytic variable}, and
the generating function of this model is denoted 
by $G(\mu)$ (the variables $\{x,y,a\}$ compose a set of parameters or weights
in the model).  The generating function $G(\mu)$ will be determined
by first finding a functional recurrence for it, using the classification of
partially directed walks illustrated in figure \ref{Fig5}.

The contributions to the generating function in figure \ref{Fig5}
are labeled by $A$ through $J$, and some 
are subtracted out while most are added in.  Accounting for horizontal and 
vertical steps, and for edge-visits, and height of the last vertex, the 
following generating functions are obtained for each of the terms:
\begin{eqnarray*}
&\hbox{A} =\Sfrac{1}{1-y\mu}; \qqq
&\hbox{B} = x\,G(\mu); \\
&\hbox{C} = \Sfrac{xy\mu}{1-y\mu}\,G(\mu);\qqq
&\hbox{D} = \Sfrac{xy/\mu}{1-y/\mu}\,G(\mu);\\
&\hbox{E} = \Sfrac{xy/\mu}{1-\y/\mu}\,G(y);\qqq
&\hbox{F} = ax\,G(0);\\
&\hbox{H} = x\,G(0);\qqq
&\hbox{J} = \Sfrac{axy\mu}{1-y\mu} \, G(0);\\
&\hbox{K} = \Sfrac{xy\mu}{1-y\mu}\, G(0).
\end{eqnarray*}
Since $G(\mu) = A\plus B\plus C\plus D\minus E\plus F\minus H\plus J\minus K$, 
this gives the following functional recurrence for $G(\mu)$:
\begin{equation}
\fl
G(\mu) = \Sfrac{1}{1-y\mu} 
+ \LB  x + \Sfrac{xy\mu}{1-y\mu}  + \Sfrac{xy}{\mu-y}\RB  G(\mu)
+ (a\minus 1)\LB  x + \Sfrac{xy\mu}{1-y\mu} \RB  G(0)
- \Sfrac{xy}{\mu-y}\, G(y) .
\label{eqnRecur}   
\end{equation}
Multiply through by $(1\minus y\mu)(y \minus \mu)$  and simplify.
The coefficient of $G(\mu)$ is the \textit{kernel}, given by
\begin{equation}
K(\mu) =
\Sfrac{y\thin\LB  1-(1\minus x\plus y^2\plus xy^2)\sfrac{\mu}{y} +\mu^2 \RB}{
(1-y\mu)(y-\mu)}  .
\end{equation}
The numerator of 
$K(\mu)$ is a quadratic with roots $\mu_0$ and $\mu_1$ such that
$\mu_1 = 1/\mu_0$ and 
\begin{equation}
\fl
\mu_0 = \Sfrac{1}{2y} \LB 
1-x+(1+x)y^2 
+\sqrt{(1-x)^2-2(1+x^2)y^2+(1+x)^2y^4}\RB  .
\end{equation}
Notice that
\begin{equation}
K(\mu) = \frac{(1-\mu\mu_0)(1-\mu\mu_1)}{(1-\mu y)(1-\mu/y)} 
\label{eqnKmu1}  
\end{equation}
so that $K(0) = 1$.

Series expansion of $\mu_0$ and $\mu_1$ shows that $\mu_1$
counts a certain set of walks.  Thus, $\mu_1$ is a ``physical root"
and it will play a key role in determine $G(\mu)$.  The substitution
$x=u(1\minus v)$ and $y=\sqrt{v}$ in 
$F=\sfrac{1}{xy}\LB  \mu_1\minus y  \RB\minus 1$ gives
\begin{equation}
F = \Sfrac{1}{2uv}
\LB  1-u(1\plus v) -
\sqrt{1-2u(1\plus v)+u^2(1\minus v)^2}
\RB  .
\end{equation}
This may be expanded in $u$ to obtain
\begin{eqnarray}
\fl
F =
u+(1\plus v)u^2+(1\plus 3v\plus v^2)u^3+& (1\plus 6v\plus 6v^2\plus v^3)u^4 \nonumber\\
&+(1\plus 10v\plus 20v^2\plus 10v^3\plus v^4)u^5 + \cdots .
\end{eqnarray}
This is the series expansion of the Narayana generating function given by
\begin{equation}
N(u,v) = \sum_{s=1}^\infty \sum_{i=1}^s
\frac{1}{s} \Bi{s}{i}\Bi{s}{i\minus 1}\, v^{i-1}u^s .
\end{equation}
That is, $F\equiv N(u,v)$ and it follows that
\begin{equation}
\mu_1 = xy(N(x,y)\plus 1)+y .
\label{eqnmu1N}   
\end{equation}

The Narayana generating function has useful properties, which will be used later.
In particular, powers of $N(u,v)$ have nice expansions:
\begin{equation}
\LB  N(u,v) \RB ^r = \sum_{s=1}^\infty \sum_{i=1}^N \frac{r}{s}
\Bi{s}{r\plus i\minus 1} \Bi{s}{i\minus 1}\, v^{i-1}u^s .
\label{Nr}   
\end{equation}

Taking $\mu=0$ in equation \Ref{eqnRecur} produces
\begin{equation}
G(y) = \Sfrac{1}{x}\LB  (1\plus x (1\minus a))\,G(0) -1 \RB  .
\end{equation}
Using equation \Ref{eqnKmu1} and the above in the recurrence
\Ref{eqnRecur} gives
\begin{equation}
\fl
\Sfrac{x(a\minus 1)}{1\minus \mu y}\,G(0)
+ \Sfrac{y}{\mu \minus  y}\,\LB  1\plus (xa\minus x\minus 1) G(0) \RB 
+ \Sfrac{1}{1\minus \mu y}
- \Sfrac{(1\minus \mu\mu_0)(1\minus \mu\mu_1)}{(1\minus \mu y)(1\minus \mu /y)}\, G(\mu) = 0.
\label{eqnfeqnf}   
\end{equation}
Substituting $\mu=\mu_1$ gives $G(0)$:
\begin{equation}
G(0) = \frac{(1\minus y^2)\mu_1}{(x(1\minus y^2)(1\minus a) - y^2)\mu_1 + y} .
\end{equation}
This is the generating function of walks ending in the adsorbing line 
(or with endpoint at height zero).  Substituting this into equation \Ref{eqnfeqnf}
gives a solution for $G(\mu)$:
\begin{equation}
G(\mu) = \frac{(1\minus y^2)\mu_1}{
(1\minus \mu \mu_1)((x(1\minus y^2)(1\minus a)\minus  y^2)\mu_1 + y)} .
\end{equation}
Recall that $K(\mu_1)=0$; this is useful in simplifying $G(\mu)$ to
\begin{equation}
G(\mu) = \frac{y\minus \mu_1}{x(1\minus \mu \mu_1)((a\minus 1)\mu_1 - ya)}.
\end{equation}
Substituting $x \to xt$, and $y\to yt$, and then doing a series expansion in $t$,
gives the series
\begin{eqnarray}
\fl
1 + (xa\plus y\mu)t +
(y^2\mu^2\plus xya\mu\plus x^2a^2\plus xy\mu)t^2 \cr
\fl\qq
+ (y^3\mu^3\plus xy^2a\mu^2\plus x^2ya^2\mu\plus 
x^3a^3\plus 2xy^2\mu^2\plus x^2ya\mu\plus x^2y\mu\plus xy^2)t^3+\cdots ,
\end{eqnarray}
of walks with $t$ the length generating variable.  Substituting $\mu_1$ using 
the Narayana generating function above gives $G(\mu)$ in terms of $N(x,y)$:
\begin{equation}
\fl
G(\mu) = \frac{N(x,y)\plus 1}{
(xa(N(x,y)\plus 1)-xN(x,y)-x-1)(xy\mu(N(x,y)\plus 1) +y\mu -1)} .
\end{equation}
Introduce $\lambda = x\plus xN(x,y)$ in equation \Ref{eqnmu1N}.  Then it 
follows that $\mu_1 = y \plus  y\lambda$, and 
$\lambda = \sfrac{1}{x}\mu_1\minus 1$.  Define $\sigma=a\minus 1$, then
it follows that 
\begin{equation}
G(\mu) = \frac{\lambda}{x(1\minus \sigma\lambda)(1-y\mu (\lambda \plus 1))} .
\end{equation}
By expanding this in a power series in  $y$, the generating function 
of walks ending in a vertex at height $h$ is
\begin{equation}
G_h(a,y) = \frac{\lambda y^h (1\plus \lambda)^h}{x(1\minus \sigma \lambda)} .
\label{eqnGH1}   
\end{equation}
Substitute $\lambda = x\plus x\thin N(x,y)$ and suppress the arguments of
$N(x,y) \equiv N$ to simplify expressions.  This gives
\begin{equation}
G_h(a,y) = \frac{(1\plus N) y^h (xN\plus x\plus 1)^h}{1-(a\minus 1)(x\plus xN)}
\label{eqnGH2}   
\end{equation}
since $\sigma=a\minus 1$, and in terms of the Narayana generating function $N(x,y)$.  
This completes the determination of the generating function $G_h(a,y)$ 
of partially directed paths ending in a vertex at height $h$.

\subsection{The generating function $G_{wh}(a,y)$ of partially directed walks of width $w$}

The generating function of walks of width $w$ and last vertex at height $h$ can be
extracted from $G_h(a,y)$ in equation \Ref{eqnGH2}.  The strategy is to expand 
$G_h(a,y)$ in powers of $\lambda = x\plus xN$, and then to use equation \Ref{Nr} 
in order to simplify the expressions.

Expanding the denominator of $G_h(a,y)$ in equation \Ref{eqnGH1} in 
$\lambda$ gives 
\begin{equation}
G_h(a,y) = \Sfrac{1}{x} \sum_{k=0}^\infty \sigma^k \lambda^{k+1} y^h
\thin (1\plus\lambda)^h .  
\end{equation}
Substituting $\lambda = x(1\plus N)$ and simplifying leaves
\begin{equation}
G_h(a,y) = \sum_{k=0}^\infty \sigma^k  x^k y^h (1\plus N)^{k+1}(1\plus x\plus xN)^h .
\end{equation}
Let $S(k)$ be the summand of $G_h(\mu)$ above.  By expanding the factors in 
$S(k)$, it follows that
\begin{equation}
S(k) = \sigma^k x^k y^h   \sum_{j=0}^h \sum_{\ell=0}^{j+k+1} 
\Bi{h}{j}\Bi{k\plus j\plus 1}{\ell}\, x^j N^\ell ,
\end{equation}
so $G_h(a,y) = \sum_{k=0}^\infty S(k)$.  Reverse the order of the summations
and allow $\ell$ to run to $\infty$ to simplify the expression.  This gives
\begin{equation}
S(k) = \sigma^k x^k y^h   \sum_{\ell=0}^\infty \sum_{j=0}^h  
\Bi{h}{j}\Bi{k\plus j\plus 1}{\ell}\, x^j N^\ell .
\end{equation}
This introduces powers of Narayana numbers which is replaced by using 
equation \Ref{Nr}.  Simplifying the resulting expression gives the following 
for $S(k)$:
\begin{eqnarray}
\fl
S(k) = y^hx^k\sigma^k \sum_{\ell=0}^\infty
\sum_{j=0}^h & \Bi{h}{j} \Bi{k\plus j\plus 1}{\ell}\,x^j \nonumber \\
& \times \LH \sum_{s=1}^\infty \sum_{i=1}^s \frac{\ell}{s}
\Bi{s}{i\plus \ell\minus 1}\Bi{s}{i\minus 1}\, \LB  \frac{x}{1\minus y^2} \RB ^s y^{2i-2} \RH .
\end{eqnarray}
The powers of $x$ in the  summand is $w=k\plus j\plus s$.  This is also the
number of horizontal steps, and so is the horizontal width $w$ of the walk.  Extract
the coefficient of $x^w$ in the above to find the generating function
of walks of width $w$ and with final vertex at height $h$.  This is, after
some simplification,
\begin{equation}
\fl
s(k,w) = \sum_{\ell=0}^\infty \sum_{s=1}^\infty \sum_{i=0}^\infty
\frac{\ell}{s} \Bi{s}{i}\Bi{s}{i\plus \ell}\Bi{h}{w\minus k\minus s}\Bi{w\minus n+1}{\ell}\,
\frac{y^{h+2i}}{(1\minus y^2)^s} .
\end{equation}
Here, the summation over $s$ is allowed to run to infinity to simplify
the expressions.  This does not introduce new terms since the binomial
coefficients introduce a natural cut-off on the sums.  It remains to 
multiplify the above by $\sigma^k$ and to sum over $k$ as well --
this gives the generating function $G_{wh}(a,y)$ of walks of width $w$ ending
in a vertex of height $h$, and with $a$ generating steps at height zero.

A minor simplification can be achieved by noting that the sum over
$\ell$ can be done.  This reduces the number of summations above to
two with the result that
\begin{equation}
\fl
s(k,w) = \sum_{s=1}^\infty \sum_{i=0}^\infty
\frac{w\plus 1\minus s}{n}
\Bi{n}{i}\Bi{h}{w\minus k\minus s} \Bi{w}{s\minus i\minus 1}\,
\frac{y^{h+2i}}{(1\minus y^2)^s} .
\end{equation}
The $s=0$ term is equal to $y^h\SBi{h}{w\minus k}$, and this should be 
inserted explicitly in the above.  Multiplying this by $\sigma^k$ and
summing over $k$ gives the generating function for walks of
width $w$ ending a vertex at height $h$ and with $a$ generating
steps at height zero.  After some simplification, this is  
\begin{equation}
\fl
G_{wh}(a,y) =  
\sum_{s=0}^w
\sum_{k=0}^w 
\sum_{i=0}^\infty \frac{w\plus 1\minus s}{w\plus i\plus 1\minus s}
\Bi{h}{w\minus k\minus s}
\Bi{n\minus 1}{i} \Bi{w}{s\minus i}\, \frac{y^{h+2i} (a\minus 1)^k  }{(1\minus y^2)^s} ,
\label{eqnGwh}   
\end{equation}
where the subsitution $\sigma = a\minus 1$ was made.

\subsection{The partition function}

The partition function of adsorbing partially directed walks of width
$w$ can be extracted from $G_{wh}(a,y)$ in equation \Ref{eqnGwh} by
determining the coefficients of $y^N$ (note that  the coefficient of $y^N$ 
in $G_{wh}(a,y)$ is the partition function of walks of length $w\plus N$).  
In other words, it remains only to expand $(1\minus y^2)^{-s}$ and to collect 
the coefficient of $y^N$.  Using the binomial theorem, putting $n=w\plus N$, 
substituting $\sigma=a\minus 1$ and simplifying, gives the partition function of 
partially directed walks with $w$ horizontal steps and length $n$:
\begin{eqnarray}
\fl
& \fl Z_{n}(w,h) = 
\sum_{i=0}^n \sum_{k=0}^w \sum_{s=0}^w 
\frac{w\plus 1\minus s}{w\plus i\plus 1\minus s} \nonumber \\
& \fl \q \times
\Bi{s\minus 1}{i}\Bi{w}{s\minus i}\Bi{h}{w\minus k\minus s}\Bi{-s}{\frac{1}{2}(w\plus h\minus n)\minus s\plus i}\,
(-1)^{(w+h-n)/2+i}\,
(a\minus 1)^k .
\end{eqnarray}
This expression can be simplified using standard binomial identities to
\begin{eqnarray}
\fl
& \fl Z_n(w,h) = 
\sum_{i=0}^n \sum_{k=0}^w \sum_{s=0}^w
\frac{w\plus 1\minus s}{w\plus i\plus 1\minus s} \nonumber \\
& \fl \q \times
\Bi{s\minus 1}{i}\Bi{w}{s\minus i}\Bi{h}{w\minus k\minus s}
\Bi{\frac{1}{2}(n\minus w\minus h)\minus i\plus s\minus 1}{s\minus 1}\,
(a\minus 1)^k .
\label{eqn35}  
\end{eqnarray}
For example, if $n=8$, $w=4$ and $h=2$, then 
\begin{equation}
Z_8(4,2) = 30 + 20\, a+ 11\, a^2 + 4\, a^3 ;
\end{equation}
the partition function of partially directed walks of length $8$ and width
$4$,  and with generating variable $a$ generating edge-visits at height 
$h=0$.  Thus, $Z_n(w,h)$ is the partition function of \textit{adsorbing partially
directed walks} of length $n$, width $w$ and endpoint at height $h$
(see figure \ref{Fig1}).

\section{The limiting entropic forces}
\label{section4}  

In this section the entropic forces arising in the limit $n\to\infty$ are examined.
Forces for finite values of $n$ will be examined in section \ref{section5} (for 
the non-interacting partially directed walk when $a=1$).

The extrinsic free energy of a walk of length $n$, width $w$, and endpoint at
height $h$ is given by
\begin{equation}
F_n(w,h) = \log Z_n(w,h)
\label{eqn39}  
\end{equation}
in terms of the partition function in equation \Ref{eqn35}.  This is 
the free energy of a walk as illustrated in figure \ref{Fig1}, confined by two
vertical walls a distance $w$ apart.  If the walls are close together (that is, when
$w$ is small), then the walk is constrained, and it loses entropy (this reduces its
extrinsic free energy).  Similarly, if $w$ is large, then the walk is stretched in
the horizontal direction, and it loses entropy as well.  The loss in entropy in
both these cases induces a restoring entropic force on the opposing walls
in figure \ref{Fig1}, moving them further apart (the repulsive regime), 
or closer together (the attractive regime).  At an intermediate distance 
the entropic force is zero, and this is the \textit{zero force point}.

\subsection{Entropic forces on confining walls}
\label{section41}   

The first priority is to determine the limiting (intrinsic) free energy
of the walk in the limit $n\to\infty$, after rescaling edges and 
lengths in the model by $\sfrac{1}{n}$ (so that edges have length 
$\sfrac{1}{n}$, and the vertical walls are a distance 
$w=\sfrac{1}{n}\lfl \alpha n \rfl$ apart).  In this limit the free energy can 
be calculated from equation \Ref{eqnGwh}, or more appropriately, 
by considering the exponentially fastest growing terms 
in equation \Ref{eqn35}.  

For general values of $a$ and $h$ it follows from equation \Ref{eqn35} that
\begin{equation*}
\fl
\Sfrac{w\plus 1 \minus s}{w\plus i\plus 1 \minus s}
\Bi{s\minus 1}{i}\Bi{w}{s\minus i}\Bi{h}{w\minus k\minus s}
\Bi{\frac{1}{2}(n\minus w\minus h)\minus i\plus s\minus 1}{s\minus 1}\,
(a\minus 1)^k \leq Z_n(w,h),
\end{equation*}
for any $n>1$, for any $a>1$, and for any $\{i,k,s\}$.  Similarly, if
$\{i_m,k_m,s_m\}$ are the values of $\{i,k,s\}$ which maximizes the
summand on the right hand side of equation \Ref{eqn35}, then
\begin{equation*}
\fl 
\Sfrac{(n\plus 1)^3(w\plus 1 \minus s_m)}{w\plus i_m\plus 1 \minus s_m}
\Bi{s_m\minus 1}{i_m}\Bi{w}{s_m\minus i_m}\Bi{h}{w\minus k_m\minus s_m}
\Bi{\frac{1}{2}(n\minus w\minus h)\minus i_m\plus s_m\minus 1}{s_m\minus 1}\,
(a\minus 1)^{k_m} 
\end{equation*}
is an upper bound on $Z_n(w,h)$.

\begin{figure}
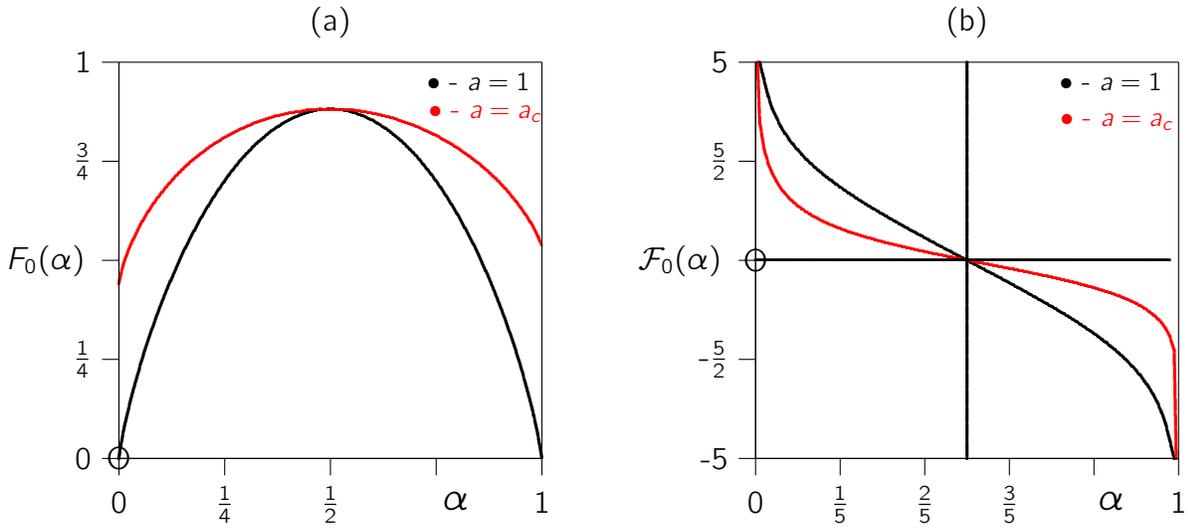

\input Fig6.tex
\caption{(a) The limiting free energy of a partially directed walk
as a function of $\alpha$.  The bottom curve is the limiting free energy
for $a=1$ and $h=0$.  The top curve is the limiting free energy for
$a=a_c$ and $h=0$.  The derivatives of these curves
are the limiting entropic force on the vertical walls, and are plotted 
as a function of $\alpha$ in (b).  The darker curve is the force curve
for $a=1$ and the lighter curve (lower on the left and higher on
the right in (b)) is the force curve for $a=a_c$.  Both curves
pass through the zero force point at $\alpha=\frac{1}{2}$.  The curve
for $a=1$ is symmetric on reflection through the point $(\frac{1}{2},0)$.}
\label{Fig6}   
\end{figure}

Rescale $(w,h)$ by putting $w=\lfl \alpha n\rfl$,
and $h= \lfl \phi n\rfl$.  The summation indices in the lower bound
are similarly scaled by $s=\lfl \epsilon n\rfl$, $i=\lfl \delta n \rfl$,
and $k = \lfl \kappa n \rfl$. Taking the power $\sfrac{1}{n}$ and then
$n\to\infty$, shows that
\begin{equation}
\fl
\sup_{\epsilon,\delta,\kappa}
\LH
\Sfrac{
\alpha^\alpha \phi^\phi
(1-\alpha-\phi-2\delta+2\epsilon)^{
(1-\alpha-\phi-2\delta+2\epsilon)/2}
(a\minus 1)^{\kappa}
}{
2^\eps \delta^\delta (\epsilon-\delta)^{2(\epsilon-\delta)}
(1-\alpha-\phi-2\delta)^{(1-\alpha-\phi-2\delta)/2}
(\phi-\alpha+\kappa+\epsilon)^{\phi-\alpha+\kappa+\epsilon}
(\alpha+\delta-\epsilon)^{\alpha+\delta-\epsilon}
(\alpha-\kappa-\epsilon)^{\alpha-\kappa-\epsilon}
}\RH
\label{eqnosup}   
\end{equation}
is a lower bound on $\lim_{n\to\infty} Z_n^{1/n}
(\lfl\alpha n\rfl, \lfl \phi n \rfl)$.  In fact, by the above this limit cannot
exceed this lower bound for any values of $(\alpha,\phi)\in [0,1]^2$.
This supremum is particularly useful in determining the free energy 
for the cases that $\phi>0$ or $a>a_c$.  Determining the free energy
for $h=0$ and $a=1$, or for $h=0$ and $a=a_c=1\plus\sfrac{1}{\sqrt{2}}$,
requires a similar approach based on equation \ref{eqn35} (but with $h=0$).
These cases are considered in sections \ref{section411} and \ref{section413} 
below.

\subsubsection{Forces when $a=1$ and $\phi=0$:}
\label{section411}
Putting $a=1$ and $h=0$ in the summand in equation \Ref{eqn35} 
reduces the partition function to a single sum.   The summand is
\begin{equation*}
\Sfrac{1}{\lfl \delta n\rfl + 1}
\Bi{\lfl\alpha n \rfl-1}{\lfl\delta n\rfl}
\Bi{\lfl\alpha n\rfl}{\lfl\alpha n\rfl - \lfl \delta n\rfl}
\Bi{\shalf(n + \lfl\alpha n\rfl) - \lfl \delta n \rfl -1}{
\lfl \delta n \rfl -1} 
\end{equation*}
where $w=\lfl \alpha n\rfl$ and $i=\lfl \delta n \rfl$.  The limiting free
energy is obtained by determining the maximum exponential rate of growth 
of this summand in the limit that $n\to\infty$.  This can be done by
using the Stirling approximation for the binomial coefficients, 
taking the power $\sfrac{1}{n}$, and then taking $n\to\infty$.
This gives
\begin{equation}
\frac{\alpha^\alpha(\alpha\minus 2\delta)^{2\delta}
(1\plus\alpha\minus2\delta)^{(1+\alpha-2\delta)/2}}{
2^\alpha \delta^{2\delta}(\alpha\minus \delta)^{2\alpha}
(1\minus\alpha\minus2\delta)^{(1-\alpha-2\delta)/2}}.
\end{equation}
The rate of growth of this is a maximum when $\delta=\delta_c$ and
\begin{equation}
\delta_c = \shalf \pm \shalf \sqrt{2\alpha^2 \minus 2\alpha \plus1}.
\label{eqn45Z}   
\end{equation}
Substituting $\delta=\delta_c$ gives the limiting free energy as a function 
of $\alpha$:
\begin{eqnarray}
\fl
F_0 (\alpha) = \alpha\log(2\alpha(1\minus\alpha))
-2\alpha\log(\sqrt{C}\plus 2\alpha\minus 1)
&+\shalf\sqrt{C}\log\LB \Sfrac{C\plus \alpha}{C\minus \alpha} \RB \nonumber
\\
&+(\sqrt{C}\minus 1) 
\log \LB \Sfrac{1\minus\sqrt{C}}{\sqrt{C}\plus2\alpha\minus 1}\RB ,
\label{eqn45}   
\end{eqnarray}
where $C=2\alpha^2\minus 2\alpha\plus 1$.  Notice that $F_0(\sfrac{1}{2}) 
= \log(1\plus\sqrt{2})$; this is not unexpected, since partially directed walks 
grow asymptotically proportionally to $(1\plus\sqrt{2})^{n+o(n)}$.  

The free energy is plotted as a function of $\alpha$ in figure \ref{Fig6}(a).
Expanding the free energy about its peak at $\alpha=\shalf$ gives
\begin{equation}
\fl
F_0(\alpha) =\log (1\plus \sqrt{2})
- 2\sqrt{2} \,(\alpha\minus \sfrac{1}{2} )^2 
-\sfrac{2}{3}\sqrt{2}\,(\alpha\minus\sfrac{1}{2})^4
+ O((\alpha\minus\sfrac{1}{2})^6) ,
\label{eqn46}   
\end{equation} 
where $O((\alpha\minus\sfrac{1}{2})^6) $ is a correction which has leading
term $\minus\sfrac{28}{15}\sqrt{2}(\alpha\minus\sfrac{1}{2})^6$.

The derivative of $\C{F}_0(\alpha) = \Sfrac{d}{d\alpha} F_0(\alpha)$ is the limiting 
entropic force on the vertical walls confining the walk.  Taking the derivative
of the approximation in equation \Ref{eqn46} to $\alpha$ gives
the force near $\alpha=\sfrac{1}{2}$:
\begin{eqnarray}
\fl
\C{F}_0(\alpha) =
- 4\sqrt{2} \,(\alpha\minus \sfrac{1}{2} ) 
-\sfrac{2^3}{3}\sqrt{2}\,(\alpha\minus\sfrac{1}{2})^3
-\sfrac{7\,\cdot\,2^3}{5}\sqrt{2}\,(\alpha\minus\sfrac{1}{2})^5
+ O((\alpha\minus\sfrac{1}{2})^7) .
\end{eqnarray}

More generally, taking the derivative of $F_0(\alpha)$ 
in equation \Ref{eqn45} gives the limiting force as a function of $\alpha$.
This is plotted in figure \ref{Fig6}(b) and simplifies to 
\begin{equation}
\C{F}_0(\alpha) = 
\log \LB
\Sfrac{2\alpha(1-\alpha)}{(C\plus 2\alpha\minus 1)^2}
\RB
+ \Sfrac{C^2\minus 2\alpha}{2C}
\log \LB
\Sfrac{(C-\alpha)(2\alpha+C-1)^2}{(C+\alpha)(1-C)^2} \RB .
\label{eqn46F}   
\end{equation}
It may be verified that $\C{F}_0(\alpha) = - \C{F}_0(1\minus \alpha)$,
so $\C{F}_0(\sfrac{1}{2}) = 0$.  That is, the limiting force
curve is symmetric on reflection through the point
$(\shalf,0)$.  For small $\alpha>0$ the above shows that
$\C{F}_0(\alpha) \sim |\log \alpha |$; this will also be shown
later (see equation \Ref{eqn63BB}).

\begin{figure}[t]
\input Fig7.tex
\caption{(a) The limiting free energy $F_\phi(\alpha,\phi)$ plotted as a function of
$\alpha$ for $\phi\in\{0,0.1,0.2,0.3,0.4,0.5\}$.  Notice that
$\alpha \in [0,1\minus \phi]$.  (b) Force curves $\C{F}_\phi(\alpha,\phi)$ as a function
of $\alpha$.  Increasing $\phi$ moves the zero force point to smaller
values of $\alpha$.  For $\phi=0$ the zero force point is located
at $\alpha_c(0) = \frac{1}{2}$. }
\label{Fig7}   
\end{figure}

\subsubsection{Forces when $a=1$ and $\phi >0$:}
\label{section412}
The limiting free energy can be determined by computing the
supremum 
\begin{eqnarray}
\fl
\sup_{\epsilon,\delta}
\LH
\Sfrac{
\alpha^\alpha \phi^\phi
(1-\alpha-\phi-2\delta+2\epsilon)^{
(1-\alpha-\phi-2\delta+2\epsilon)/2}
}{2^\eps
\delta^\delta (\epsilon-\delta)^{2(\epsilon-\delta)}
(1-\alpha-\phi-2\delta)^{(1-\alpha-\phi-2\delta)/2}
(\phi-\alpha+\epsilon)^{\phi-\alpha+\epsilon}
(\alpha+\delta-\epsilon)^{\alpha+\delta-\epsilon}
(\alpha-\epsilon)^{\alpha-\epsilon}
}\RH ,
\end{eqnarray}
which is found after putting $\kappa=0$ and by taking $a\to 1^-$
in equation \Ref{eqnosup}.  The supremum is realised when $\epsilon
=\eps_s$, and $\delta=\delta_s$,  where $\eps_s$ and $\delta_s$
are given by (note that $0 \leq \alpha\plus\phi \leq 1$)
\begin{eqnarray}
\eps_s &= \frac{
\alpha (\alpha \minus \alpha^2 \plus  
\sqrt{(2\alpha^2\minus 2\alpha\plus 1)\phi^2 \minus \phi^4} )
}{(\alpha\plus \phi)(1\plus \phi\minus\alpha )
}\\
\delta_s &=\frac{\alpha}{2\phi}
\frac{(1\minus\alpha\minus \phi)(\phi\plus \phi^2 \minus  
\sqrt{(2\alpha^2\minus 2\alpha\plus 1)\phi^2 \minus \phi^4} )
}{(\alpha\plus\phi)(1\plus \phi \minus\alpha )}
\end{eqnarray}
Substituting these values to determine the supremum (using
Maple \cite{Maple}) gives a very lengthy expression for the free energy 
$F_\phi (\alpha,\phi)$ as a function of  $(\alpha,\phi)$.  It is plotted 
against $\alpha$ for  various values of $\phi$ in figure \ref{Fig7}(a).

The force $\C{F}_\phi (\alpha,\phi)$ is the (partial) derivative of 
$F_\phi (\alpha,\phi)$ to $\alpha$.  This is  plotted against $\alpha$ for 
various values of $\phi$ in figure \ref{Fig7}(b). If $\phi=0$ then 
the force vanishes when $\alpha=\shalf$, so that $\C{F}_\phi(\shalf,0) = 0$ 
(where $\phi=0$).  This is the zero force point for the case $\phi=0$.  
The zero force point is generally located at $a_c(\phi)$, so $a_c(0)=\shalf$.
If $\phi$ increases, then the zero force point moves to smaller values of
$\alpha$.  By expanding the force in $\phi$ and $\alpha$,
the location of the zero force point as a function of $\phi$ can be
estimated by determining the first few terms in a series expansion
in $\phi$.  To order $\phi^6$ this is
\begin{equation}
\fl
\alpha_c (\phi) = \shalf - \sfrac{1}{4}\sqrt{2}\,\phi^2
+\sfrac{1}{4}(3\sqrt{2} \minus 2)\,\phi^4 +
\sfrac{1}{48}(144\minus 149\sqrt{2})\,\phi^6 + O(\phi^8) .
\end{equation}
This estimate is surprisingly accurate.  For example,
$a_c(0.1) = 0.4964\ldots$ and the estimate is
$a_c(0.1) \approx 0.4965\ldots$, and $a_c(0.3)=0.4669\ldots$
while the estimate is $a_c(0.3)\approx 0.4717\ldots$.  The estimate
for $\phi=0.5$ is $a_c(0.5)\approx 0.4249\ldots$ as opposed
to the exact value $a_c(0.5)=0.40095\ldots$.

\subsubsection{Forces when $a=a_c$ and $\phi=0$:}
\label{section413}
Put $a=a_c = 1\plus\sfrac{1}{\sqrt{2}}$ in summand in
equation \Ref{eqn35} and bound the partition
function in a way similar to the way it was done in section \Ref{section41}.
Take the power $\sfrac{1}{n}$, and let $n\to\infty$. This shows that
the limiting free energy can be determined by computing critical 
values of $(\epsilon, \delta, \kappa)$ to find the supremum
\begin{eqnarray}
\fl
\sup_{\epsilon,\delta,\kappa}
\LH
\Sfrac{
\alpha^\alpha 
(1-\alpha-2\delta+2\epsilon)^{
(1-\alpha-2\delta+2\epsilon)/2}
(\sfrac{1}{\sqrt{2}})^{\kappa}
}{
\delta^\delta (\epsilon-\delta)^{2(\epsilon-\delta)}
(1-\alpha-2\delta)^{(1-\alpha-2\delta)/2}
(\kappa+\epsilon-\alpha)^{\kappa+\epsilon-\alpha}
(\alpha+\delta-\epsilon)^{\alpha+\delta-\epsilon}
(\alpha-\kappa-\epsilon)^{\alpha-\kappa-\epsilon}
}\RH \! .
\end{eqnarray}
The critical values of $(\epsilon,\delta,\kappa)$ in the above are
\begin{eqnarray*}
\fl
\epsilon_s &= \Sfrac{1}{4}(2\minus\sqrt{2})
\LB
1\plus \alpha\plus
\sqrt{4\sqrt{2}\,\alpha(1\minus \alpha)
\plus (1\plus \alpha)^2}
\RB ; \\
\fl
\delta_s & = \Sfrac{1}{2}(2\minus\sqrt{2})\LB 1\minus \alpha \RB ; \\
\fl
\kappa_s &= \Sfrac{1}{4}(2\minus\sqrt{2})
\LB
(1\plus\sqrt{2})^2\alpha \minus 1 \minus
 \sqrt{4\sqrt{2}\,\alpha(1\minus \alpha)
\plus (1\plus\alpha)^2}
\RB .
\end{eqnarray*}
Substituting these, simplifying, and then taking the logarithm
gives a lengthy expression for the limiting free energy $F_c(\alpha)$ as a function 
of $\alpha$.  Expanding this in $\alpha$ about $\alpha = \shalf$ gives, after simplification,
\begin{eqnarray}
\fl
F_c (\alpha) &=\log (1\plus\sqrt{2}) - \sfrac{4}{7}
(2\sqrt{2}\minus 1)\,(\alpha\minus \sfrac{1}{2})^2 
+\sfrac{8}{7^3}(\sqrt{2}\minus 1)\,(\alpha\minus\sfrac{1}{2})^3 
\nonumber \\
\fl
&\qqq -\sfrac{4}{3\,\cdot\,7^5}
(1\plus\sqrt{2})^2(1\minus2\sqrt{2})^6(1\plus 8\sqrt{2})\,
(\alpha\minus \sfrac{1}{2})^4
+ O((\alpha\minus\sfrac{1}{2})^5) .
\end{eqnarray}
Notice that $F_c(\alpha)$ is not symmetric
about $\alpha=\sfrac{1}{2}$, unlike $F_0(\alpha)$ (the free energy
when $\phi=0$ and $a=1$).  Taking the derivative
of $F_c(\alpha)$ to $\alpha$ gives the force near $\alpha=\sfrac{1}{2}$:
\begin{eqnarray}
\fl
\C{F}_c(\alpha) &=
\sfrac{2^3}{7}(2\sqrt{2}\minus 1)\,(\alpha\minus \sfrac{1}{2}) 
+\sfrac{3\,\cdot 2^3}{7^3}(\sqrt{2}\minus 1)\,(\alpha\minus\sfrac{1}{2})^2 
\nonumber \\
\fl
&\qqq - \sfrac{2^4}{3\,\cdot\,7^5}
(1\plus\sqrt{2})^2(1\minus2\sqrt{2})^6(1\plus 8\sqrt{2})\,
(\alpha\minus \sfrac{1}{2})^3
+ O((\alpha\minus\sfrac{1}{2})^4) .
\end{eqnarray}

The full expression of the free energy $F_c(\alpha)$ is a lengthy expression 
determined using Maple \cite{Maple} and it is plotted in figure \ref{Fig6}(a). 
Taking its derivative to $\alpha$ shows that the force vanishes when
$\alpha=\shalf$, consistent with the series expansion above.  
Thus, $\C{F}_c(\shalf) = 0$.  The force curve is plotted in figure \ref{Fig6}(b).

\subsubsection{Forces when $a=a_c$ and $\phi>0$:}
\label{section414}
If $a=a_c$ and $\phi>0$, then the critical values of $(\epsilon,\delta,\kappa)$
in equation \Ref{eqnosup} are given by
\begin{eqnarray*}
\fl
\epsilon_s &= \Sfrac{1}{4}(2\minus\sqrt{2})
\LB
1\plus \alpha\minus \phi\plus
\sqrt{4\sqrt{2}\,\alpha(1\minus \alpha \minus \phi)
\plus (1\plus \alpha\minus \phi)^2}
\RB\! ; \\
\fl
\delta_s & = \Sfrac{1}{2}(2\minus\sqrt{2})\LB 1\minus \alpha \minus \phi \RB \! ; \\
\fl
\kappa_s &= \Sfrac{1}{4}(2\minus\sqrt{2})
\LB
(1\plus\sqrt{2})^2\alpha \minus 3\phi \minus 1 \minus
 \sqrt{4\sqrt{2}\,\alpha(1\minus \alpha \minus \phi)
\plus (1\plus\alpha\minus \phi)^2}
\RB \! ,
\end{eqnarray*}
Substituting these into equation \Ref{eqnosup}, simplifying, and then taking 
the logarithm, gives a lengthy expression for the limiting free energy $F_c(\alpha,\phi)$
as a function of $\alpha$ and $\phi$.  The entropic forces are again given by
the partial derivative $\C{F}_c(\alpha,\phi) = \sfrac{\partial}{\partial\alpha} 
F_c(\alpha,\phi)$. 

\begin{figure}
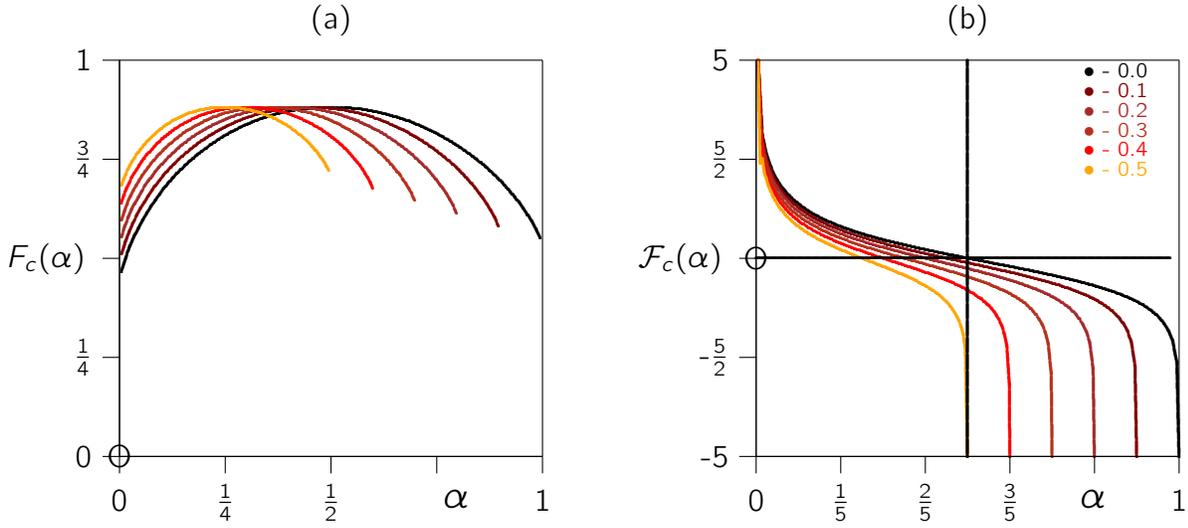

\input Fig8.tex
\caption{(a) The limiting free energy $F_c(\alpha)$ plotted as a function of
$\alpha$ for $\phi\in\{0,0.1,0.2,0.3,0.4,0.5\}$.  Notice that
$\alpha \in [0,1\minus \phi]$.  (b) Force curves $\C{F}_c (\alpha)$ as a function
of $\alpha$.  Increasing $\phi$ moves the zero force point to smaller
values of $\alpha$, and this point is located at exactly
$\frac{1}{2}\minus\frac{1}{2}\phi$.}
\label{Fig8}   
\end{figure}

In figure \ref{Fig8} the limiting free energy and forces are plotted
as a function of $\alpha$ for fixed $\phi$.     With increasing $\phi$ the
zero force point $\alpha_c(\phi)$ moves to smaller values of $\alpha$. 
The location of this point may be approximated by expanding the force 
in $\phi$ and in $\alpha$, and then solving for $\alpha$ as a function of 
$\phi$.  The result is
\begin{equation}
\alpha_c(\phi) = \sfrac{1}{2} - \sfrac{1}{2}\phi + O(\phi^2) .
\end{equation}
Numerical analysis of the model suggests that the correction $O(\phi^2)$ is 
zero, so $\alpha_c(\phi)$ is equal to $\sfrac{1}{2} \minus \sfrac{1}{2}\phi$.  
This is verified by substitution and then simplification of the expression for the
force.  That is, the zero force point is located exactly at
\begin{equation}
\alpha_c(\phi) = \sfrac{1}{2} - \sfrac{1}{2}\phi.
\label{eqn60}   
\end{equation}
if $a=a_c$.

\subsubsection{Forces when $a>a_c$ and $\phi>0$:}
\label{section415}  
The adsorbed phase of the model is obtained if $a>a_c$.  In this case the
critical values of $(\epsilon,\delta,\kappa)$ in equation \Ref{eqnosup}
are given by
\begin{eqnarray*}
\fl
\epsilon_s &= \Sfrac{1}{4a}
\LB
1\plus \alpha\minus \phi\plus
\sqrt{8(a\minus 1)\alpha(1\minus \alpha \minus \phi)
\plus (1\plus \alpha\minus \phi)^2}
\RB ; \\
\fl
\delta_s & = \Sfrac{1}{2a}\LB 1\minus \alpha \minus \phi \RB ; \\
\fl
\kappa_s &= \Sfrac{1}{4a}
\LB
(4a\minus 1)\alpha \minus 3\phi \minus 1 \minus
 \sqrt{8(a\minus 1)\alpha(1\minus \alpha \minus \phi)
\plus (1\plus\alpha\minus \phi)^2}
\RB ,
\end{eqnarray*}
provided that  $a>a_c$. 

Notice that
$\eps_s \plus \kappa_s = \alpha \minus \Sfrac{1}{a}\phi  \leq \alpha$
as required in equation \Ref{eqnosup}, provided that
$a\alpha \geq \phi$.  Similarly, 
$\phi \minus \alpha \plus \eps_s \plus \kappa_s
 = \phi (1 \minus \sfrac{1}{a}) \geq 0$, provided that $a\geq 1$.
In addition, $\eps_s \geq \delta_s$ 
and  $1\minus \alpha \minus \phi \minus  2\delta_s
\geq 0$ if both $a\geq 1$ and $\alpha \plus \phi \leq 1$. 

Substitution of the above critical values for $(\eps,\delta,\kappa)$
and taking the logarithm gives the limiting (intensive) free energy 
$F(\alpha,\phi)$.  This expression for $F(\alpha,\phi)$ is lengthy
(it was determined using Maple \cite{Maple}) and will not be reproduced 
here.  

\begin{figure}
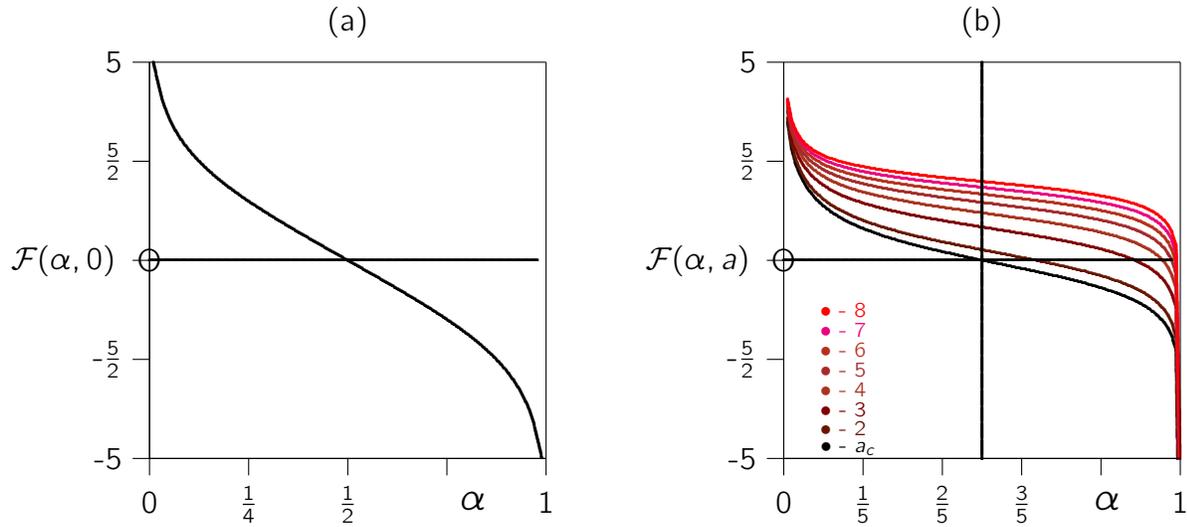

\input Fig9.tex
\caption{(a) The entropic force curve for $a=1$ and $\phi=0$ is
recovered from the force $\C{F}(\alpha,0)$.  This curve is identitical
to the force curve $\C{F}_0$ for $a=1$, plotted in figure
\ref{Fig6}(b).  (b)  The force curves $\C{F}(\alpha,a)$ for
increasing values of $a$.  If $a=a_c$, then the zero force point is
at $\alpha=\frac{1}{2}$.  Increasing $a$ moves the zero force point
to the right and flattens the force curve, compared to the force
curve $\C{F}_0$ plotted in (a).}
\label{Fig9}   
\end{figure}

Taking the derivative of $F(\alpha,\phi)$ to $\alpha$ gives the
limiting entropic force $\C{F}(\alpha,\phi)$.  Putting
both $\phi=0$ and $a=1$ simplifies $\C{F}(\alpha,\phi)$ to equation
\Ref{eqn46F}, and this force is plotted in figure \ref{Fig9}(a).

In figure \ref{Fig9}(b) the forces for values of $\phi=0$ and $a\geq a_c$ and 
for $\phi=0$ are plotted.  The expressions for these forces are lengthy,
and were obtained using Maple \cite{Maple}.  At the critical
adsorption point $a_c = 1\plus\sfrac{1}{\sqrt{2}}$ the zero force point
is located at $\alpha=\shalf$, unchanged from its location when $a=1$.
 
Increasing the value of $a$ gives force curves which are progressively 
higher in the graph (and so larger if repulsive, and less attractive, if 
attractive), while the zero force point moves towards larger values of 
$\alpha$.  If $\phi=0$, then the location of the zero force point may be
estimated by expanding the force in $\alpha$ and solving for the
zero force point.  This gives the following asymptotic expression
for the location of the zero-force point:  If $a\minus a_c$ is small,
then the zero force point is located at 
\begin{equation}
\alpha_c(a) =  \shalf + \shalf (a \minus a_c)
- \sfrac{1}{4}(\sqrt{2}\minus 1)(a\minus a_c)^2 + O((a\minus a_c)^3).
\label{eqn61}   
\end{equation}
For example, if $a = a_c \plus 0.1$, then 
$\alpha_c(a_c\plus 0.1) = 0.5488\ldots$ while the approximation
gives $\alpha_c(a_c\plus 0.1) \approx 0.544\ldots$.  Similarly,
 $\alpha_c(a_c\plus 0.05) = 0.5247\ldots$ while the approximation
gives $\alpha_c(a_c\plus 0.05) \approx 0.524\ldots$.

For large values of $a$ an asymptotic expression for the zero force point
may be found.  This is given by
\begin{equation}
\alpha_c(a) = 1 - \Sfrac{2}{(a-1)^3} + \Sfrac{4}{(a-1)^4} - \Sfrac{10}{(a-1)^5}
+ O( (a\minus 1)^{-6}) .
\label{eqn62}   
\end{equation}
This becomes very accurate for large values of $a$.  Numerically
$\alpha_c(4.0) = 0.9559\ldots$ while the above approximation gives
$\alpha_c(4.0)\approx 0.934\ldots$.  Similarly, 
$\alpha_c(6.0) = 0.9884\ldots$ while the approximation gives
$\alpha_c(6.0)\approx 0.987\ldots$.

\section{Finite size asymptotics for $a=1$ and $h=0$}
\label{section5}  

In this section an asymptotic expression for the partition function
in equation \Ref{eqn35} (this is for the special case that $a=1$ and $h=0$) is 
determined.  Notice that if $h=0$ then only terms with $w-k-s=0$ survive, 
and if $a=1$, then $k=0$, so that $w=s$.  This reduces the triple summation 
in equation \Ref{eqn35} to a single summation:
\begin{equation}
Z_n(w,0) = \sum_{i=0}^w \frac{1}{i\plus 1} 
\Bi{w\minus 1}{i} \Bi{w}{w\minus i} \Bi{\shalf(n\plus w)\minus i\minus 1}{w\minus 1}
\label{eqn46B}  
\end{equation}
This partition function will be approximated by an integral, and the integral will be
approximated by a saddle point formula.  The first step is to find a good
approximation for the summand.

Thus, consider the summand in equation \Ref{eqn46B} and denote it by $S$.  
Its binomial coefficients will be approximated by using the Stirling 
approximation for the factorial:
\begin{equation}
n! = \sqrt{2\pi n} \thin n^n e^{-n} (1-\Sfrac{1}{12n}+\Sfrac{1}{288n^2} + O(\sfrac{1}{n^3})).
\end{equation}
Take the logarithm of the summand in equation \Ref{eqn46} and expand the binomial
coefficients into factorials. This gives
\begin{eqnarray}
\fl & \fl\log S =  \log (i\plus 1) - \log (w\minus i) 
                     + \log ((\sfrac{1}{2}(n\plus w)\minus i\minus 1)!) \nonumber \\
 &\fl\q - \log ((\sfrac{1}{2}(n\plus w)\minus i\minus 1)!) 
                     - 2\log ((w\minus i\minus 1)!) + \log(w!) - 2\log((i\plus 1)!).
\end{eqnarray}  
Substitute the factorial by the Stirling approximation above, and simplify.
This gives an expression approximating the logarithm of the summand.

In order to implement a saddle point approximation, it is necessary to find
the location of the saddlepoint in the summand.  Thus, in the simplified approximation
of $\log S$ above, substitute $n=\sfrac{1}{\eps}$, $i=\sfrac{\delta}{\eps}$ and
$w=\sfrac{\alpha}{\eps}$.  Expand the resulting expression in $\eps$ and collect
the leading term (which is $O(\eps^{-1})$).  This is a lengthy expression in $\delta$
and $\alpha$, and the saddle point is located by taking its derivative with respect to
$\delta$ and solving for the stationary point.  This gives
$\delta = \sfrac{1}{2} \pm \sfrac{1}{2} \sqrt{2\alpha^2\minus 2\alpha\plus 1}$
as the possible locations of the saddle point(s) (see for example
equation \Ref{eqn45Z}). Notice that the choice of the
plus sign before the square root gives $\delta>\alpha$ -- so that $i>w$ in the
summand.  This is outside the range of $i$ in equation \Ref{eqn46}, with the
result that the minus sign before the square root is the correct choice of the
sign.  Thus, the saddle point is located at asymptotic values of $i$ in the summand 
where $i= \lfl \delta_s n\rfl$, where
\begin{equation}
\delta_s = \Sfrac{1}{2} - \Sfrac{1}{2} \sqrt{2\alpha^2\minus 2\alpha\plus 1} .
\end{equation}
 
Numerical work shows that the width in the peak about $\delta_s$ is proportional
to $\sqrt{n}$, thus the summation in equation \Ref{eqn46} will be 
approximated by an integral by putting $n=\sfrac{1}{\eps^2}$, $w=\sfrac{\alpha}{\eps^2}$
and $i=\sfrac{\delta_s}{\eps^2} + \sfrac{\delta}{\eps}$ in the logarithm of the
summand.  The result is expanded in a Laurent series in $\eps$; this gives leading terms
of order $O(\sfrac{1}{\eps^2})$ and $O(\sfrac{1}{\eps})$.  The series is truncated to
$O(\eps)$, then exponentiated and simplified, before it is integrated
over $\delta$.  The result will be a saddle point approximation
to the partition function $Z_n(w,0)$ with $w=\lfl \alpha n \rfl$, once the
substitution $\eps = \sfrac{1}{\sqrt{n}}$ is made.  After significant 
symbolic computations calculations
using Maple \cite{Maple}, the approximation to $Z_n(w,0)$ for $a=1$ is
\begin{eqnarray}
Z_n(\lfl \alpha n \rfl,0) 
&\sim 
\frac{1-\alpha+\alpha^2+\sqrt{C}}{\pi \thin n^2 \alpha(\alpha\minus 1)^2
 \sqrt{(1\minus \alpha)\sqrt{C}+C}} 
\label{eqn51}  
\\
&\q\times \LB \frac{1-3\alpha+3\alpha^2+(1\minus 2\alpha)\sqrt{C}
}{\alpha^\alpha(1\minus \alpha)^\alpha}\RB^n
\LB \frac{\sqrt{C}+\alpha}{\sqrt{C}-\alpha} \RB^{n/2} \! , \nonumber
\end{eqnarray}
where $C=2\alpha^2\minus 2\alpha\plus 1$.

\begin{figure}
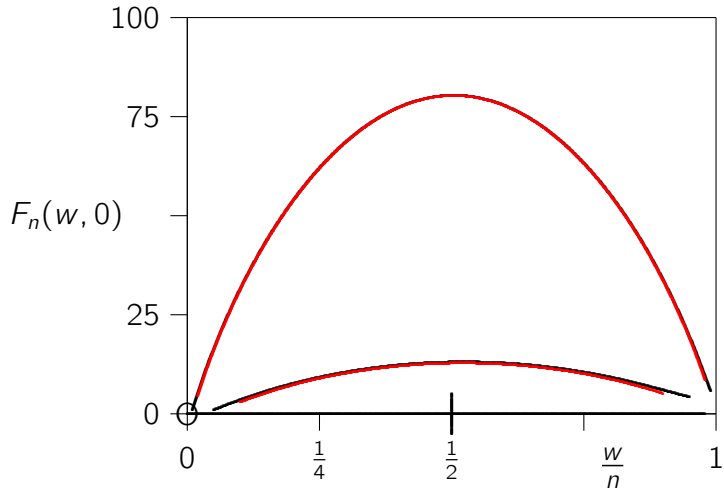

\input Fig10.tex
\caption{The exact and approximate free energies $F_n(w,0)$ for $n=20$ 
(bottom curves) and $n=100$ (top curves).  The approximate expression is obtained by
$F_n(w,0) = \log Z_n(w,0)$ where $Z_n(w,0)$ is approximated by equation 
\Ref{eqn51}.  On this scale there is little differences between the exact
curve and its approximation.}
\label{Fig10}   
\end{figure}

The approximation for $Z_n$ above also approximates the extensive free
energy, given by $F_n(\lfl \alpha n \rfl,0) = \log Z_n(\lfl \alpha n \rfl,0)$ according to
equation \Ref{eqn39}.   In figure \ref{Fig10} the exact extensive free energy,
and its approximation (by the logarithm of the right hand side of
equation \Ref{eqn51}), are plotted on the same graph as a function of
$\sfrac{w}{n}$ for $n=20$ and $n=100$.  On the scale of this plot, the approximation is indistinguishable
from the exact curve.

An asymptotic formula for the entropic force is obtained by taking
the derivative of $\log Z_n(\lfl \alpha \rfl, 0)$ in equation \Ref{eqn51}
to $\alpha$ while recalling that $w=\lfl \alpha n \rfl$.  In figure \ref{Fig11}
the exact force curve, and the asymptotic approximation to it, are plotted
on the same graph for $n=200$.  This shows that the forces are very small for
$\alpha=\sfrac{1}{2}$.  In fact, taking the logarithm and then the derivative of
the right hand side of equation \Ref{eqn51}, gives a complicated expression
which simplifies to $1\plus\sqrt{2}$ if $\alpha=\sfrac{1}{2}$.  That is, there
is no dependence on $n$ at this point, and it may be verified, using 
symbolic computations \cite{Maple}, that the
coefficient of $n$ vanishes at this point.  That is, the asymptotic expression 
of the finite size force is of the form  
\begin{equation}
\C{F}_n (\alpha) \sim A_\alpha + B_\alpha n
\end{equation}
and $A_\alpha$ and $B_\alpha$ are functions of $\alpha$ such that
$A_{1/2} = 1\plus\sqrt{2}$ and $B_{1/2}=0$.  Thus the force does not
vanish at $\alpha=\sfrac{1}{2}$, but is, instead, a constant value, independent
of $n$, where $\C{F}_n (\sfrac{1}{2}) = 1\plus\sqrt{2}$.  Since $\C{F}_n(\alpha)$
increases with $n$ for $\alpha\not=\sfrac{1}{2}$, this indicates that
\begin{equation}
\Sfrac{\C{F}_n (\sfrac{1}{2})}{\C{F}_n(\alpha)} \to 0,\:\hbox{as $n\to\infty$
and $\alpha\not=\sfrac{1}{2}$}.
\end{equation}

One may similarly use equation \Ref{eqn51} to determine an asymptotic expression
for the entropic force on the two vertical walls in figure \ref{Fig1}.  This
may be done by taking the derivative of the free energy to $\alpha$.
For $\alpha$ small the result is
\begin{equation}
\C{F}_n (\alpha) \sim
-\Sfrac{1}{\alpha} + (n\minus\sfrac{1}{2}) - n \log \alpha +O(n\alpha).
\label{eqn63BB}   
\end{equation}
Fixing $w=\lfl n\alpha\rfl$ and then taking $\alpha\to 0^+$ shows that
$\C{F}_n(\alpha) \sim n |\log \alpha| \to \infty$ for small values of $n$.
Moreover, $\sfrac{1}{n} \C{F}_n(\alpha) \sim |\log \alpha |$ (see
section \ref{section411}).

\section{Numerical calculation of the entropic forces}
\label{section6}

\subsection{Entropic force on the vertical wall}

For finite values of $n$ and $w$ the entropic force may be
defined by the finite difference
\begin{equation}
\C{F}_n(w,h) = \Sfrac{1}{2}\LB F_n(w\plus 2,h) - F_n(w,h)\RB 
\label{eqn40}  
\end{equation}
of the extensive free energy. The difference between the free energies
for widths $w\plus 2$ and $w$ are used in order 
to avoid parity effects in the model; these parity effects are due to fact
that the underlying square lattice is a bipartite graph.

\begin{figure}
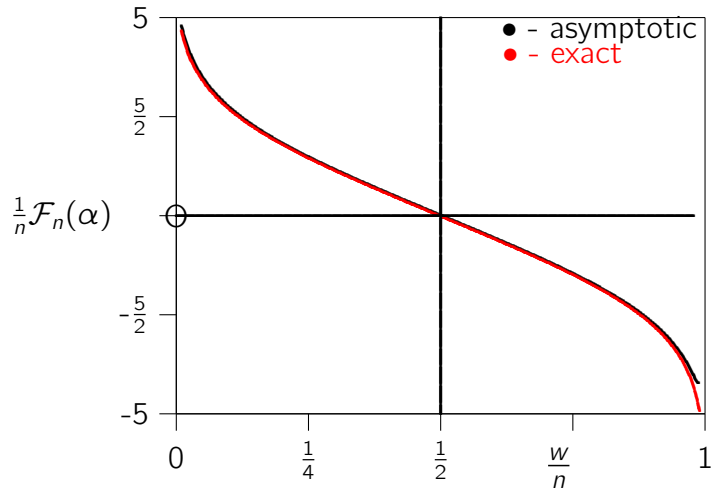

\input Fig11.tex
\caption{ The normalised force curve $\frac{1}{n}\C{F}_n(\alpha)$
for $n=200$ and $a=1$ (see figure \ref{Fig2}(b)) plotted as a function of $w$.  
There are two curves, namely, the exact force for $n=200$, computed 
from equation \Ref{eqn40}, and the asymptotic approximation
to $\C{F}_n(\alpha)$ for $n=200$. The curves pass very close to $\frac{1}{2}$ when 
$\C{F}_n=0$. }
\label{Fig11}   
\end{figure}

\subsubsection{The case $a=1$:}  In this case there is no attraction between
 the walk and the adsorbing line,  and the model is in its desorbed phase.

In figure \ref{Fig2}(a) the entropic force $\C{F}_{20}(w,0)$ is interpolated
on discrete points as a function of the width $w$ for walks of length $n=20$ 
and endpoint fixed at height $h=0$.  As expected, the force between the
confining walls is large positive (repulsive) for small values of $w$, and large
negative (attractive) for large values of $w$.  At a critical value of $w$,
namely $w\approx 10$, the force is zero, neither attractive nor repulsive.

Forces of paths of different lengths $n$ can be compared by scaling $w$ by $n$
in figure \ref{Fig2}(a).  In figure \ref{Fig2}(b) the interpolated
force curves for $n\in\{20,40,60,80,100\}$ are plotted against 
$\alpha = \sfrac{w}{n}$.  These curves collapse to a single curve, namely a
universal force curve which is the limiting force in the limit $n\to\infty$ with
length rescaled by $n$.

Note that the force curves in figure \ref{Fig2} pass very close to the
point $(w,n) = (\sfrac{1}{2} n,n)$ when $\C{F}_n(w,0)=0$.  That is, the force 
is approximately zero when $w=\sfrac{1}{2} n$, or when the confining
walls are $w=\sfrac{1}{2}n$ apart.  This occurs when the walk gives 
$\sfrac{1}{2}n$ horizontal steps, and this is consistent with the location
of the zero force point at $\alpha=\sfrac{1}{2}$ determined in section
\ref{section411}.

The convergence of the finite size forces to the limiting force curve
is quick, this can be seen, for example, in figure \ref{Fig11},
where the limiting force, and the exact force for $n=200$, are plotted
on the same graph.  These curves are indistinguishable on this scale.

\begin{figure}
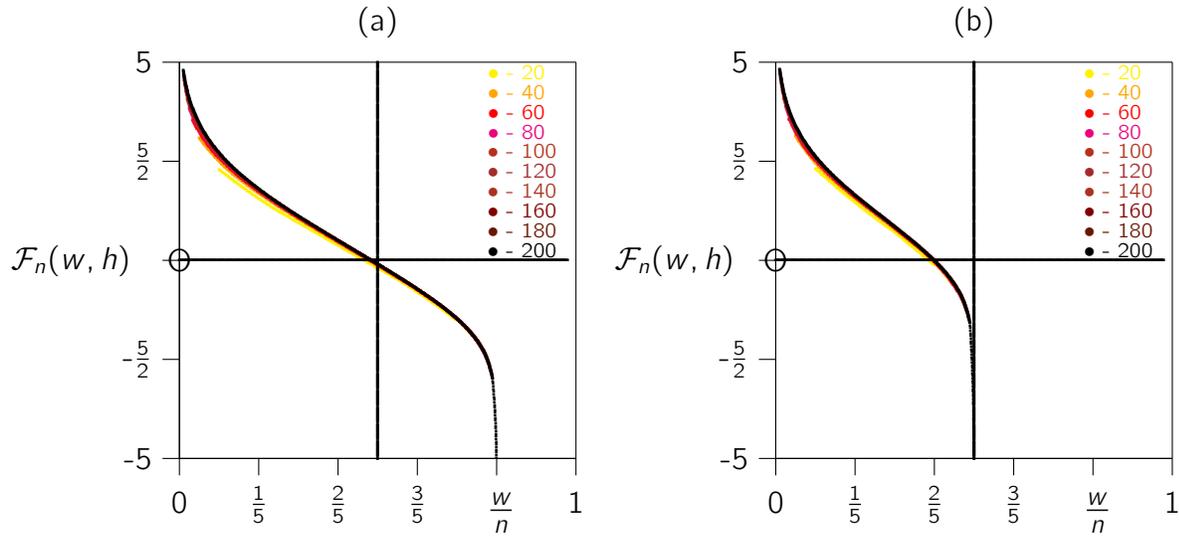

\input Fig12.tex
\caption{The entropic forces exerted by a partially directed walk between two vertical walls.  
The width $w$ of the walk (or distance between the confining walls) is rescaled
by $n$ (the length of the walk).  (a) The entropic forces for $n$ between 
$20$ and $200$ plotted for walks with $a=1$ and height of endpoint 
$h=\frac{1}{5}n$.  Notice that this distorts the curves when $\frac{w}{n}$ is larger
than about $0.5$.  (b) Similar to (a), but now with $h=\frac{1}{2}n$. }
\label{Fig12}   
\end{figure}

In figure \ref{Fig12} the effects of the height of the endpoint on the
induced forces are displayed.  In figure \ref{Fig12}(a) the force curves are
plotted for $h=\lfl \phi n \rfl$ with $\phi=\sfrac{1}{5}$.  Since $w$ is rescaled by 
$n$, the curves again collapse to an underlying force curve.  
For small values of $w$ the force curves are not changed
significantly from those in figure \ref{Fig2}(b).  In figure \ref{Fig12}(b)
similar data are presented, but now with $h=\sfrac{1}{2}n$.

\begin{figure}
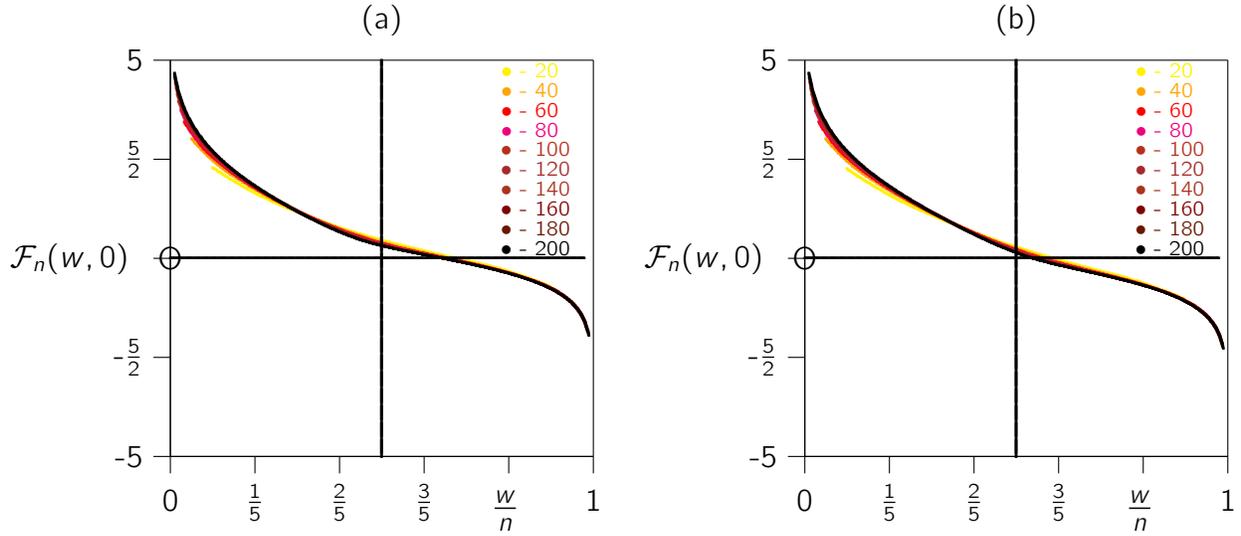

\input Fig13.tex
\caption{The entropic forces exerted by a partially directed walk between two vertical walls.  
The width $w$ of the walk (or distance between the confining walls) is rescaled
by $n$ (the length of the walk) and the height of the endpoint of the
walk is at $h=0$.  (a) The entropic forces plotted for $a=2$. 
The force curves are shifted in the positive direction compared to 
the results in figure \ref{Fig2}(b).  This shows that the forces are repulsive
for larger values of $w$, and, when they become attractive (negative),
then they are weakened.  (b) The entropic forces plotted for $a=a_c$ 
(the critical adsorption point in the model).}
\label{Fig13}   
\end{figure}

Finally, the effects of the adsorption activity $a$ is examined in figures
\ref{Fig13}(a) and \ref{Fig13}(b).  For $a$ larger than $1$ the walks are 
attracted to the adsorbing boundary $\partial \mathL^2_+$, and they
adsorb at $a=a_c=1\plus \sfrac{1}{\sqrt{2}}$.  In figure \ref{Fig13}(a)
the forces are plotted for $a=2>a_c$.  The data shows a longer range
of repulsive forces, and for large $w$, when the forces are attractive,
the magnitude of the forces are reduced.  This distortion of the
forces can be seen even in the case that $a=a_c$, as shown in
figure \ref{Fig13}(b), although here it is more subtle.  There are, however,
an apparent reduction in the magnitude of the attractive forces when
$w$ approaches $n$.

\subsection{The pressure on the adsorbing wall}

Let $u=\edge{(w,0)}{(w\plus 1,0)}$ be an edge in the adsorbing boundary
$\partial\mathL^2_+$ a distance $w$ from the origin.   The pressure
$\Pi_n(w)$ on $u$ is the change in extensive free energy if $u$ is a blocked
so that walks cannot pass through it.  That is, if $f_n$ is the free energy
of walks of length $n$, and $f_n(\o{u})$ is the free energy of walks avoiding
the edge $u$, then the pressure on $u$ is given by
\begin{equation}
\Pi_n(w) = f_n - f_n(\o{u}) .
\label{eqn41}   
\end{equation}
The (extensive) free energies $f_n$ and $f_n(\o{u})$ are determined by
the corresponding partition functions, and the partition function of walks
avoiding $u$ will be found by first computing the partition function
of walks passing through $u$.

\begin{figure}
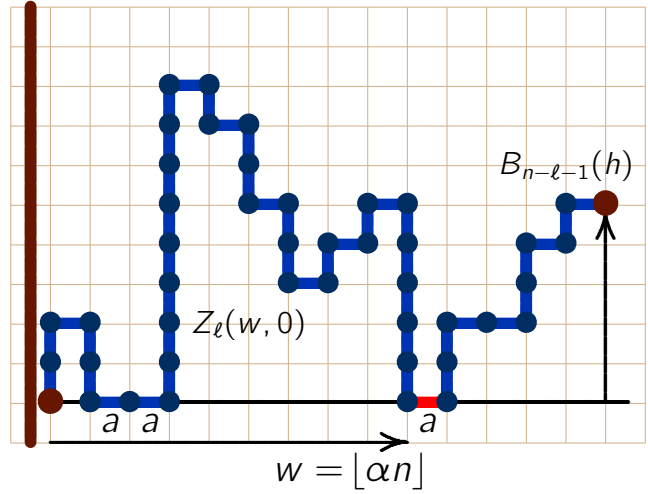

\centering
\input Fig14.tex
\caption{A partially directed walk passing through an edge in the adsorbing
boundary a distance $w$ from the origin may be calculated by
concatenating a bargraph path of width $w$ with a partially directed walk
as shown.  The bargraph path and partially directed walk are joined
by inserting an edge-visit $u=\edge{(w,0)}{(w\plus 1,0)}$ between 
the endpoint of the bargraph path and the first vertex in the 
partially directed walk.  Since $u$ is an edge-visit, it
carries weight $a$, similar to other edge-visits in the model.}
\label{Fig14}   
\end{figure}

As in equation \Ref{eqn35}, denote the partition function of partially directed
walks of length $n$ from the origin in $\mathL^2_+$, with edge-visits to
$\partial\mathL^2_+$ weighted by $a$, of horizontal width $w$ and with
endpoint at height $h$, by $Z_n(w,h)$.  This partition function can be used to
determine the partition function of walks passing through an edge $u$
as illustrated in figure \ref{Fig14}.

Let $b_n(v,h)$ be the number of partially directed walks of length $n$
in $\mathL^2_+$, from the origin, making $v$ edge-visits to $\partial\mathL^2_+$,
and with final vertex at vertical height $h$.  The partition function of this walk
is given by
\begin{equation}
B_n(h) = \sum_{v=0}^n b_n(v,h)\thin a^v ,
\end{equation}
and it is implicitly a function of the adsorbing activity $a$.
If $\mu$ is introduced as the generating variable conjugate to $h$, and
$t$ is conjugate to $n$, then the generating function of this walk is given by
\begin{equation}
\fl
B(t,\mu,a) = 
\frac{t
(1\minus \mu t)
\LB 1-(1\plus 2\mu)t+t^2+t^3 - \sqrt{D} \RB
}{
\LB (1\plus \mu\plus \mu^2)t - (1\plus t^2\plus t^3)\mu \RB
\LB (1\plus a)t^2 + (1\minus a)(1\minus t\plus t^3\minus \sqrt{D} ) \RB
} '
\label{eqnB}   
\end{equation}
Summing $B_n(h)$ over $h$ gives the partition
function of walks ending at any height, namely $P_n = \sum_{h=0}^n B_n(h)$,
and this is also implicitly a function of $a$.  

Notice that in terms of $Z_n(w,h)$,
\begin{equation}
B_n(h) = \sum_{w=0}^n Z_n(w,h),\ande P_n = \sum_{w=0}^n \sum_{h=0}^{n-w} Z_n(w,h) . \label{eqn43}  
\end{equation}
The partition function of adsorbing bargraph paths is $B_n(0)$.

The partition function of walks passing through the edge $u$ may be obtained 
concatenating a bargraph path of width $w$, with a partially directed walk ending
in a vertex at height $h$.  That is, concatenate a walk generated by
$Z_\ell (w,0)$ with a path generated by $B_{n-\ell-1}(h)$, as illustrated in 
figure \ref{Fig14}, and then sum over $\ell$.  That is, the partition function 
of walks passing through the edge $u$ and ending in a vertex at height $h$ is 
\begin{equation}
Y^{(1)}_n(h) =  a\sum_{\ell=w}^{n-1} Z_\ell(w,0)\,B_{n-\ell-1}(h) .
\label{eqn42}  
\end{equation}
Similarly, for walks ending in a vertex at any height, the partition function is
given by
\begin{equation}
Y^{(2)}_n = a \sum_{\ell=w}^{n-1} Z_\ell(w,0)\,P_{n-\ell-1}.
\end{equation}
Note that the extra factor $a$ in these expressions is the weight of the edge
$u$, which is not accounted for in the partition function otherwise.  See
figure \ref{Fig14} for more explanation.

The pressure on the edge $u$ by walks of length $n$ can be found from equation
\Ref{eqn41} by determining $f_n$ and $f_n(\o{u})$.  In particular, for walks
ending a vertex at height $h$, $f_n \equiv \log B_n(h)$, and $f_n(\o{u}) \equiv
\log (B_n(h) \minus Y^{(1)}_n(h))$ since the partition function of walks ending 
in a vertex at height $h$ which avoids the edge $u$ is $B_n(h) \minus Y_n^{(1)}(h)$.  
Thus the pressure on $u$ by walks ending in a vertex at height $h$ is given by
\begin{equation}
\fl
\Pi^{(1)}_n(w,h) =  \log B_n(h) - \log \LB B_n(h) \minus Y_n^{(1)}(h) \RB 
= - \log\LB 1 \minus \Sfrac{Y_n^{(1)}(h)}{B_n(h)} \RB .
\label{eqn46C}  
\end{equation}

\begin{figure}
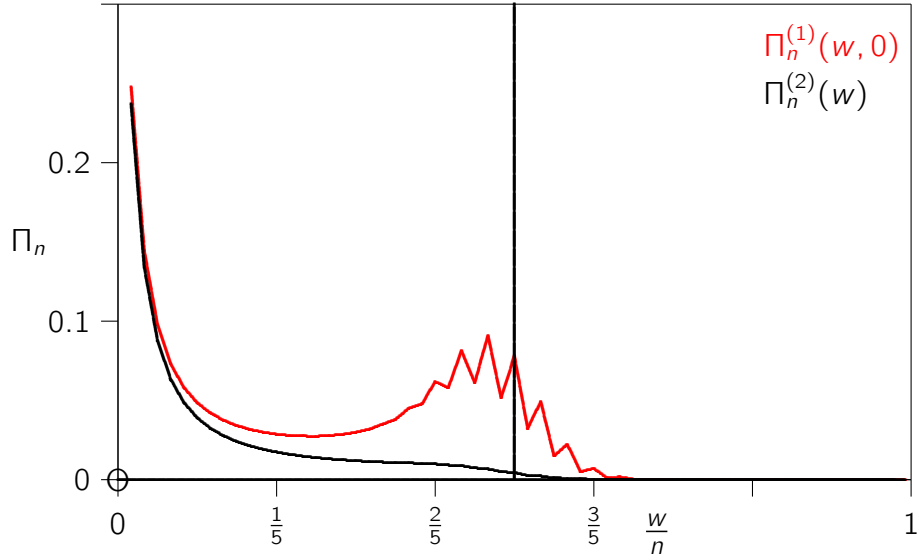

\input Fig15.tex
\caption{The pressures $\Pi^{(1)}(w,0)$ and $\Pi^{(2)}(w)$ as a function
of $w$.  Notice that $w$ is normalised by $n$ and that the pressures
are scaled by multiplication with $n$.   In these simulations $a=1$ and
$n=100$. The pressure of a bargraph path $\Pi^{(1)}(w,0)$ on an edge 
$u$ in the adsorbing line at first decreases quickly with increasing $w$,
but then goes through a secondary, peak strongly modified by a parity
effect, at about $w=\frac{1}{2} n$.  For $w$ approaching $n$ the
pressure decreases to zero.  The pressure of partially directed walks
$\Pi^{(2)}(w)$ with unrestricted endpoint decreases monotonically
with increasing $w$ to small values for $w>\frac{1}{2} n$.}
\label{Fig15}   
\end{figure}

\begin{figure}
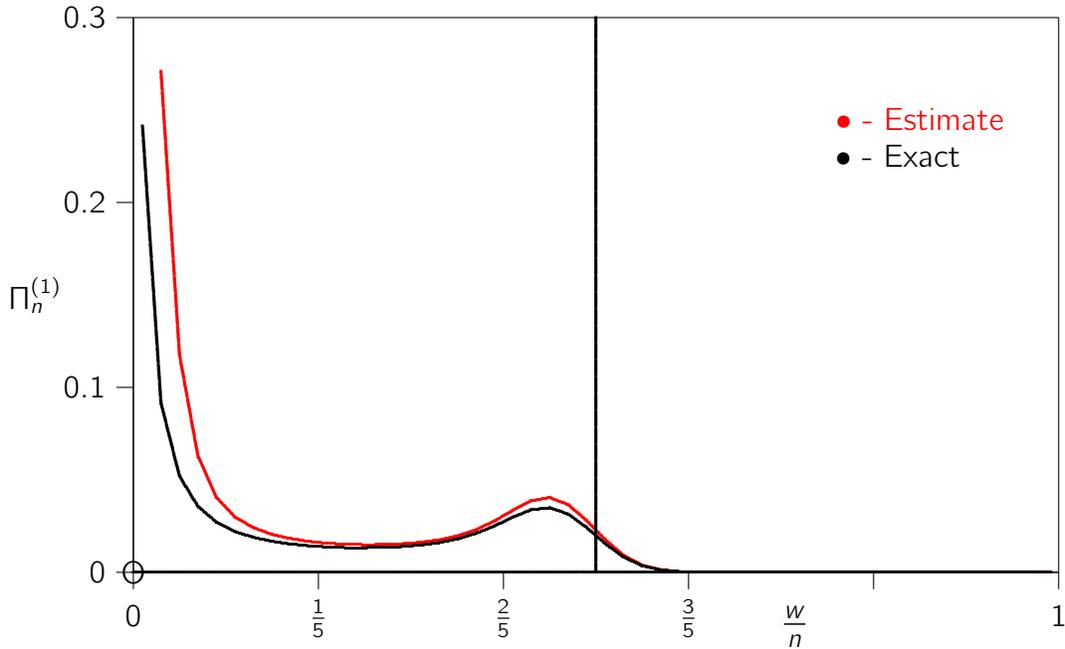

\input Fig16.tex
\caption{The pressure $\Pi^{(1)}(w,0)$ of bargraph paths of length $n=100$ 
and $a=1$ on $\partial\mathL^2_+$ interpolated on odd values of $w$. The pressure 
is large close to the origin, declines steeply with increasing $w$, and then 
passes through a secondary peak before decaying to zero.  The estimated pressure
were computed by using the approximation in equations \Ref{eqn52}
and \Ref{eqn53} in equation \Ref{eqn46C}.}
\label{Fig16}   
\end{figure}

Next, the pressure $\Pi_n^{(1)}(w,0)$ in equation \Ref{eqn46C} can be
approximated by approximations of $Y_n^{(1)}(0)$ (see equation \Ref{eqn42})
and the bargraph partition function $B_n(0)$.   $Y_n^{(1)}(0)$ is approximated
by using the asymptotic approximation \Ref{eqn51} for $Z_n(w,0)$,
and for $B_n(0)$ the asymptotic approximation
\begin{equation}
\fl
B_n(0) \sim \Sfrac{1}{\sqrt{\pi\thin n^3}} \LB 1\plus \sqrt{2} \RB^{n+3/2}
\LB 1 - \Sfrac{36+21\sqrt{2}}{16\thin n} + \Sfrac{1745+1260\sqrt{2}}{256\thin n^2}
+ O(n^{-3}) \RB ,
\label{eqn52}  
\end{equation}
derived in reference \cite{JvRR02}, can be used.  Substitute these
in equation \Ref{eqn42}, and approximate the summation by an integral (notice
that the summation in equation \Ref{eqn42} includes values of $\ell$ close
to zero, and again close to $n$, where the approximations in equations
\Ref{eqn51} and \Ref{eqn52} are poor).   Thus, by restricting the summation 
to $w>4$, the resulting approximation will be
\begin{equation}
Y_n^{(1)}(0) \simeq a \sum_{\ell=w+1}^{n-5} Z_\ell(w,0)\,B_{n-\ell-1}(0) .
\label{eqn53}  
\end{equation}
That is, using this approximation, and the approximation for $B_n(0)$ above,
$\Pi_n^{(1)}(w,0)$ may be approximated using equation \Ref{eqn46C}. 
This approximation is interpolated on odd values of $w$ for $n=100$ in
figure \ref{Fig16} (there are strong parity effects, as seen in figure \ref{Fig15}).

Similarly, the pressure of walk ending in a vertex at any height on the
edge $u$ is given by
\begin{equation}
\fl
\Pi^{(2)}_n(w) =  \log P_n - \log \LB P_n \minus Y_n^{(2)} \RB 
= - \log \LB 1 \minus \Sfrac{Y_n^{(2)}}{P_n} \RB\!\! ,
\end{equation}
since $f_n \equiv P_n$ in equation \Ref{eqn41} (where $P_n$ is the partition
function of partially directed walks of length $n$ ending at any height), and
$f_n(\o{u}) \equiv P_n \minus Y_n^{(2)}$ is the partition function of
walks of length $n$ ending at any height, and avoiding the edge $u$.

The pressures $\Pi^{(1)}_n(w,h)$ and $\Pi^{(2)}_n(w)$ can be computed
for small values of $n$ (say $n\leq 100$) and involves quadruple summations
(three in equation \Ref{eqn35} and another in equation \Ref{eqn43}). These
expressions were coded into a C program in order to explore the pressures
$\Pi^{(1)}$ and $\Pi^{(2)}$.  Numerical simulations for $a=1$ show a strong parity 
dependence on $w$ (the distance of the edge $u$ from the origin).
For example, in figure \ref{Fig15} the pressures are displayed as a function
of $w$ (normalised by $n$) for $n=100$.  The pressure $\Pi_n^{(1)}(w,0)$ decreases
quickly with increasing $w$, but shows a secondary peak for $w$ approaching
$\sfrac{1}{2} n$, with a strong correction due to parity effects.  The 
pressure $\Pi_n^{(2)}(w)$, for walks with endpoint at any height, 
decreases monotonically to zero with increasing $w$.    Parity effects for
small values of $n$ remains visible in the results; thus, in order to suppress
these, pressures were evaluated only for odd values of $w$, in what follows
below.

\begin{figure}
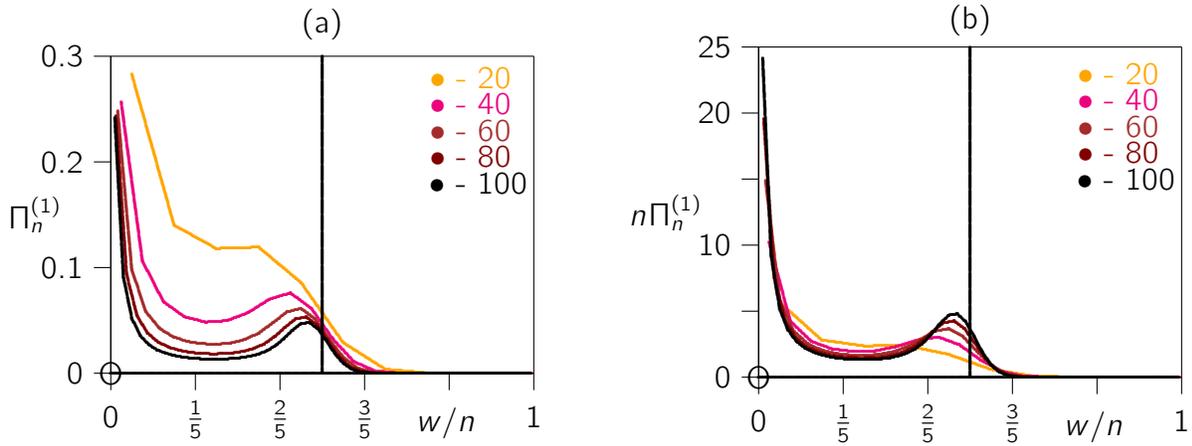

\input Fig17.tex
\caption{The pressures $\Pi^{(1)}(w,0)$ as a function
of $w$ with $a=1$. (a)  The pressures for $n$ in $\{20,40,60,80,100\}$ against
$\frac{w}{n}$.  With increasing $n$ the pressure decreases, but it shows a secondary
peak when $w$ approaches $\frac{1}{2}n$. (b) Since the length of a
rescaled edge decreases proportional to $\frac{1}{n}$ with increasing $n$, the
pressures in (a) can be rescaled by plotting $n\Pi^{(1)}_n$ against $\frac{w}{n}$.  This
brings the data in (a) closer together and suggests an underlying
limiting rescaled pressure.  Note that the data in these graphs are interpolated 
on odd values of $w$.}
\label{Fig17}   
\end{figure}

In figure \ref{Fig17}(a) the pressure $\Pi^{(1)}_n(w,0)$ for bargraph
walks are interpolated on odd values of $w$ and plotted against $\sfrac{w}{n}$
for $n\in\{20,40,60,80,100\}$.   For any fixed $n$ the pressure decreases
with increasing $w$, but there is a secondary peak in the pressure when 
$w$ approaches $\sfrac{1}{2}n$.   The appearance of this secondary peak may
be understood by noting that the endpoint of bargraph paths in 
$\partial\mathL^2_+$ would cluster around the expected horizontal 
width of the path, and so exerts pressure on edges in this vicinity.

It is also noticeable that for fixed values of $\sfrac{w}{n}$ the 
pressure declines with increasing $n$. This follows because, with increasing $n$,
the likelihood of a walk passing through the edge $u=\edge{w}{(w\plus 1)}$ 
decreases.  Since the rescaling of $w$ by $n$ effectively gives edges 
of length $\sfrac{1}{n}$, a multiplication of $\Pi^{(1)}_n(w,0)$ by $n$ should 
rescale the pressure curves to compensate; there may even be a 
limiting (non-zero) pressure curve in the limit as $n\to\infty$ in this model.  
This possibility is explored in figure \ref{Fig17}(b), and while there remains 
some spread in the curves, they do cluster nicely together.

\begin{figure}
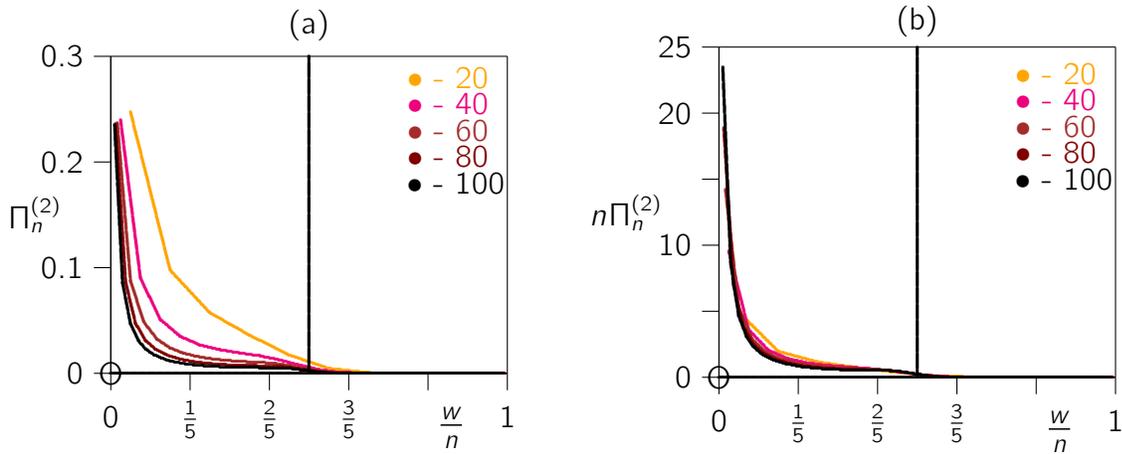

\input Fig18.tex
\caption{The pressures $\Pi^{(2)}(w)$ as a function
of $w$ with $a=1$. (a)  The pressures for $n$ in $\{20,40,60,80,100\}$ as
interpolated as a function of $\frac{w}{n}$.  With increasing $n$ the pressure 
decreases monotonically. (b) Since the length of a rescaled edge decreases 
proportional to $\frac{1}{n}$ with increasing $n$, the pressures in (a) can be rescaled 
by plotting $n\Pi^{(2)}_n$ against $\frac{w}{n}$.  This collapses the data in (a) closer 
together and suggests an underlying limiting rescaled pressure.  Note that 
the data in these graphs are interpolated on odd values of $w$.}
\label{Fig18}   
\end{figure}

In figure \ref{Fig18}(a) the data for the pressure $\Pi_n^{(2)}(w)$ is 
interpolated on $\sfrac{w}{n}$.  Since the endpoint of the walk is unconstrained
in this case, the pressure declines to zero with increasing $w$ without the
presence of a secondary peak (see figure \ref{Fig17}(a)).  These data can
also be rescaled by plotting $n\thin\Pi_n^{(2)}$ in figure \ref{Fig18}(b) against
$\sfrac{w}{n}$.  This exposes, as in figure \ref{Fig17}(b), an apparent
limiting pressure curve. 

\begin{figure}
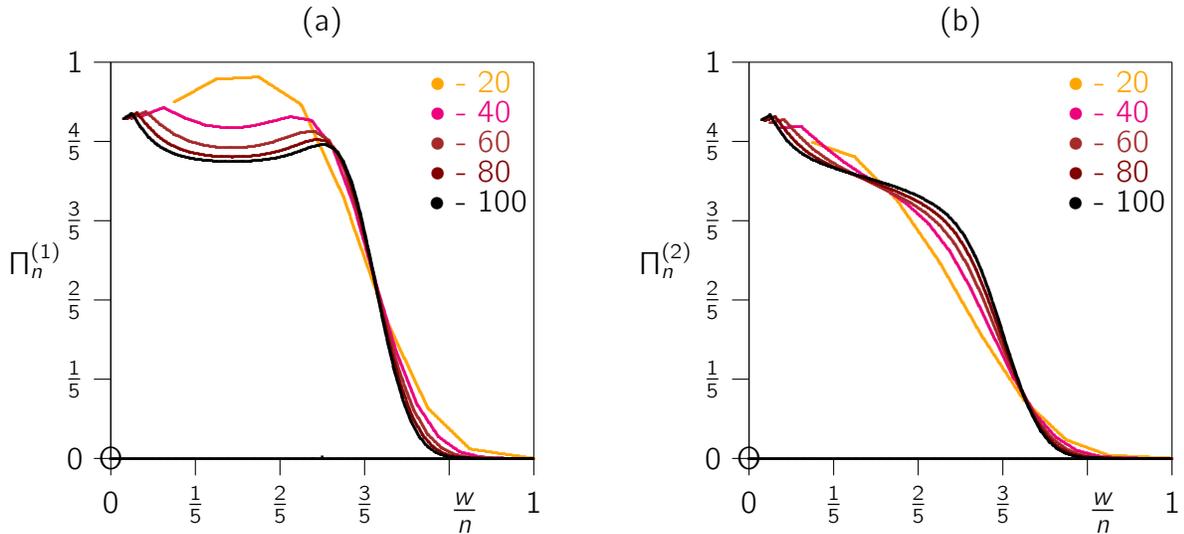

\input Fig19.tex
\caption{The pressures $\Pi^{(1)}(w,0)$  (for partially directed walks with endpoint
at height $h=0$ and with $a=2$) and $\Pi^{(2)}(w)$ (for partially directed walks with 
endpoint at any height $h$ and with $a=2$) as a function of $w$. 
(a)  The pressures forpartially directed paths walks (or bargraph paths)
with endpoint at height $h=0$ for $n$ in $\{20,40,60,80,100\}$ as a 
interpolated as a function of $\frac{w}{n}$.  For $w$ less than about
$\frac{1}{2}n$ the pressures are large, and while slowly decreasing,
may approach a non-zero constant as $n\to\infty$.  For $w$ large the
pressure approaches zero.  Note that there is no-rescaling of $\Pi^{(1)}(w,0)$
by $n$ in this figure, unlike the graph displayed in figure \ref{Fig17}(b).
(b)  The pressures for walks with endpoint at any height, for
$n$ in $\{20,40,60,80,100\}$, as interpolated as a function of 
$\frac{w}{n}$.  These pressures were also not rescaled by $n$, unlike
the data in figure \ref{Fig18}(b).  For small values of $w$ the pressures
are large and non-vanishing, but for large values of $w$ the pressures
approaches zero.  Note that the data in this graphs are interpolated 
on odd values of $w$. }
\label{Fig19}   
\end{figure}

In figure \ref{Fig19} the pressures are displayed for $a=2$; that is, when
the walk is adsorbed onto $\partial\mathL^2_+$.   Comparison to the data
in figures \ref{Fig17}(a) and \ref{Fig18}(a) shows that the pressures here are
larger, and moreover, for small values of $w$, do not decrease markedly with
increasing $n$.  In fact, for values of $w$ less than about $\sfrac{1}{2}n$ both
the data for $\Pi_n^{(1)}$ and $\Pi_n^{(2)}$ suggest that the pressures converges
to a non-zero value as $n$ becomes very large. 
For large values of $w$, the pressures approach zero.

\section{Conclusions}
\label{section7}

In this paper a two dimensional
partially directed walk model was used to examine the forces
and pressures of a linear polymer near an adsorbing wall.  We determined the
generating function of the model, and used it to compute the entropic forces
it induces between two vertical confining walls, and also to determine the
pressure exerted on the adsorbing wall.

Similar lattice models of the entropic forces in polymers have been examined
elsewhere.  For example, in reference \cite{Guttmann09} a self-interacting 
self-avoiding walk model of pulled walks was examined in $\mathL^2$ using
exact enumeration data.  In the constant distance ensemble the force-extension
curves were determined for walks below, at, and above the $\theta$-temperature.
The results (see for example figure 9 in \cite{Guttmann09}) show that the
force is attractive (pulling the end-points of the walks together) if the distance
between the endpoints is a non-zero
constant fraction of $n$ (the length of the walk).  This is
not surprising, since the natural length scale in this model is $n^\nu$ (where
$\nu < 1$ is the metric exponent of the self-avoiding walk).  Thus, there is no
zero force point in this model.

It is possible to prove that there is no zero force point in a model of
self-avoiding walks confined to a slit
of fixed width $w$.  This may be done as follows.  
A \textit{bridge} is a self-avoiding walk from the origin, with first
step in the $x$-direction, never to return to the $x=0$ plane, and with 
last vertex having maximal $x$-coordinate
(see, for example, reference \cite{MS93} for a definition).  
If a bridge has $x$-span equal to $w$, then it has \textit{width} $w$.
If the number of bridges of length $n$ and width $w$ is denoted $\beta_n(w)$, 
then the growth constant $\mu_w$ is known to exist
(see, for example, reference \cite{JvROW06}), and is given by
\begin{equation}
\log \mu_w = \lim_{n\to\infty} \sfrac{1}{n} \log \beta_n(w) .
\end{equation}
The force on the vertical walls is given by the discrete derivative
\begin{equation}
F_w = \log \mu_w - \log \mu_{w-1}.
\end{equation}
It is known that $\mu_{w-1} < \mu_w <\mu$, where $\mu$ is the growth
constant of the self-avoiding walk \cite{JvROW06}.  This shows that
$F_w>0$ for all values of $w$, and there is no zero force point in this
model.  Moreover, taking $w\to\infty$ shows that $\lim_{w\to\infty} F_w=0$,
since $\mu_w\to \mu$ as $w\to\infty$ \cite{JvROW06}.

An alternative model, with length in the horizontal direction scaled by $n$
is obtained by choosing $w=\lfl \alpha n \rfl$ in the above model.
The partition function of this model if defined by 
$Z_n(\alpha) = \beta_n(\lfl \alpha n \rfl)$.
The (extensive) free energy of this model is given by
is $F_n(\alpha) = \log Z_n(\alpha)$.
The microcanonical density conjugate to the width of bridges in this model
is defined by the limit
\begin{equation}
P(\alpha) = \lim_{n\to\infty} \LB \beta_n(\lfl \alpha n \rfl) \RB^{1/n} .
\end{equation}
It can be shown that this limit exists  (see reference \cite{JvR15}),
and is a log-concave function of $\alpha\in [0,1]$, and so is
differentiable for almost every $\alpha\in(0,1)$.  

Define the free energy of \textit{pulled bridges} in the \textit{constant force 
ensemble} by
\begin{equation}
\C{F}(x) = \lim_{n\to\infty} \sfrac{1}{n} \log \sum_{w=0}^n
\beta_n(w)\, x^w
\end{equation}
where $x = e^{f}$ (and $f$ is the horizontal pulling force on the endpoint
of the bridge).  This limit can also be shown to exist (see reference \cite{JvR15}
for more details).   The microcanonical density function $P(\alpha)$ is
related to the Legendre transform of $\C{F}(x)$, namely
\begin{equation}
\log P(\alpha) = \inf_{x>0} \{ \C{F}(x) - \alpha \log x \} .
\end{equation} 
It is known that $\C{F}(x)$ is a convex function of $\log x$ and that
\begin{equation}
\C{F}(x)  \cases{
= \log \mu, & \hbox{if $x\leq 1$}; \\
> \log \mu, & \hbox{if $x>1$}.
}
\end{equation}
For proofs of these facts, see references \cite{Beaton2015,JvRW16}.
This implies that, (1) the right derivative of $P(\alpha)$ at 
$\alpha=0$ is equal to $0$, and (2) $P(\alpha)$ is strictly decreasing
for $\alpha\in (0,1]$.  

The limiting force on the vertical walls in this model is given by
\begin{equation}
F_\alpha = \sfrac{d}{d\alpha}\, \log P(\alpha) .
\end{equation}
By the properties of $P(\alpha)$, this shows that $F_\alpha$ is a
strictly negative force for all values of $\alpha \in (0,1)$ (implying
that it pulls the vertical walls together).  Moreover, $F_0=0$
and $F_\alpha$ is a (strictly) decreasing function of $\alpha>0$,
showing that the magnitude of the pulling force increases as the endpoints
of the bridge are taken further apart. That is, there are
no zero force points in this model as well.  

In contrast to the results for the self-avoiding walk discussed above, 
our results show that the forces between confining walls may be either
attractive (if the endpoints of the walk are far apart), or repulsive (if the
endpoints are close together).  Generally the strength of the repulsive
forces increases if the walk adsorbs on the adsorbing wall, as seen, for example,
in figure \ref{Fig13}(b).  

A particular interesting result is the location of the
zero force point in the models.  For bargraph paths (when $h=0$ and $a=1$
in section \ref{section411}) this point is located at $\alpha_c(0)=\shalf$, so that
the force vanishes when the horizontal extent of the path is one-half its
total length.  The location of this point does not move with increasing
attraction into the adsorbing line in the desorbed phase, as seen in 
section \ref{section413}, where at the critical point $a_c$, the zero force
point is still located at $\alpha = \shalf$.   In other words, for all desorbed
bargraph paths, the zero force point is located at $\alpha = \shalf$ (this
may be verified by using the full expression for the free energy
$F(\alpha,\phi)$ in section \ref{section415} (and by putting $\phi=0$ therein).

For values of $a\geq a_c$ and $\phi>0$ the location
of the zero force point becomes more interesting.  In particular, for $a=a_c$
the location of this point is a function of $\phi$ given in equation \Ref{eqn60},
and approximate locations for the zero force point in the adsorbed phase
(where $a>a_c$) are given for $\phi=0$ in equations \Ref{eqn61} and
\Ref{eqn62}.

We also succeeded in developing an asymptotic formula for the partition
function of the partially directed walk in the case that $a=1$ and $h=0$ (these 
are bargraph paths).  This result was used to verify our results, and it shows 
excellent agreement.  

We examined the pressure of the walk on the adsorbing wall -- 
this is always positive, and is sharply peaked near the origin
(where the walk is tethered to the adsorbing wall).  Rescaling the pressure
(see for example figure \ref{Fig17}) shows a secondary peak at the average
position of the freely moving endpoint of the walk.

\vspace{0.5cm}
\noindent{\bf Acknowledgements:} EJJvR acknowledges financial support 
from NSERC (Canada) in the form of a Discovery Grant.  

\vspace{0.5cm}
\noindent{\bf References}
\bibliographystyle{plain}
\bibliography{References}

\end{document}